\documentclass{nature}

\newcommand{\gpmlpt}{GPM\,J1839\ensuremath{-}10}             
\newcommand{\gpmlptT}{$P_\text{orb} = 31482.4 \pm 0.2$ s}                   
\newcommand{\gpmlptPa}{$P_1 = 1318.1957 \pm 0.0002$ s}            
\newcommand{\gpmlptPb}{$P_2 = 1265.2197 \pm 0.0002$ s}            
\newcommand{\gpmlpti}{$i = 100.1 \pm 0.6^\circ $}            
\newcommand{\gpmlptalpha}{$\alpha = 52.1 \pm 0.4^\circ $}    
\newcommand{\gpmlptphiz}{$\phi_0 = 179.29 \pm 0.04^\circ $}  
\newcommand{\gpmlptzeta}{$\zeta = 61 \pm 2^\circ $}          
\newcommand{\gpmlptWspin}{$W_\text{spin} = 65 \pm 3^\circ $} 
\newcommand{\gpmlptWorb}{$W_\text{orb} = 70 \pm 1^\circ $}     
\newcommand{\jlpt}{J1912\ensuremath{-}44}

\newcommand{\jlpti}{$i \in [53,  65]^\circ$}
\newcommand{\jlptalpha}{$\alpha \in [ 42,  70]^\circ$}
\newcommand{\jlptphiz}{$\phi_0 \in [-48, -17]^\circ$}
\newcommand{\jlptzeta}{$\zeta \in [ 39,  73]^\circ$}
\newcommand{\jlptWspin}{$W_\text{spin} \in [20,  60]^\circ$}
\newcommand{\jlptWorb}{$W_\text{orb} \in [67,  90]^\circ$}

\newcommand{\revised}[2][]{#2}

\usepackage{aas_macros}
\usepackage{amsmath}
\usepackage{amssymb}
\usepackage{hyperref}
\usepackage{lineno}
\usepackage{float}
\usepackage{booktabs}
\usepackage{enumitem}

\hypersetup{colorlinks,citecolor=blue,linkcolor=blue,urlcolor=blue}

\DeclareUnicodeCharacter{2212}{-}

\title{A binary model of long period radio transients and white dwarf pulsars}

\author{
Csan\'{a}d Horv\'{a}th$^{1}$,
Nanda Rea$^{2,3}$,
Natasha Hurley-Walker$^{1}$,
Samuel J. McSweeney$^{1}$,
Richard A. Perley$^{4}$,
Emil Lenc$^{5}$
}

\usepackage{graphicx}
\makeatletter
\let\saved@includegraphics\includegraphics
\AtBeginDocument{\let\includegraphics\saved@includegraphics}
\renewenvironment{figure}{\@float{figure}}{\end@float}
\makeatother
      
\begin{document}

\newcommand{\lum}{erg\,s\ensuremath{^{-1}}}
\newcommand{\flux}{erg\,cm\ensuremath{^{-2}}\,s\ensuremath{^{-1}}}

\def\arcsec{\mbox{$^{\prime\prime}$}}
\def\nh{\hbox{$N_{\rm H}$}}

\maketitle

\begin{affiliations}
 \item International Centre for Radio Astronomy Research, Curtin University, Kent St, Bentley WA 6102, Australia
\item Institute of Space Sciences (ICE, CSIC), Campus UAB, Carrer de Can Magrans s/n, 08193, Barcelona, Spain
\item Institut d’Estudis Espacials de Catalunya (IEEC), Esteve Terradas 1, RDIT Building, 08860, Castelldefels, Spain
 \item National Radio Astronomy Observatory, Socorro, NM 87801, USA
 \item Australia Telescope National Facility, CSIRO, Space \& Astronomy, Epping, New South Wales, Australia
\end{affiliations}

\begin{abstract}

Long-period radio transients (LPTs) represent a recently uncovered class of Galactic radio sources exhibiting minute-to-hour periodicities and highly polarised pulses of second-to-minute duration. Their phenomenology does not fit exactly in any other class, although it might resemble that of radio magnetars or white dwarf (WD) radio emitting binary systems. Notably, two LPTs with confirmed multi-wavelength counterparts have been identified as WD -- M dwarf binaries. Meanwhile, systems such as AR\,Scorpii and \jlpt{} exhibit short-period pulsations in hrs-tight orbits, with polarised radio emission proposed to be generated by the interaction of the WD magnetosphere with the low-mass companion wind. 

\revised{Here, we investigate the longest-lived LPT known, \gpmlpt{}, demonstrating that it has a $\sim$8.75\,hr orbital period. We show that its radio pulses can be modelled in the same geometric framework as WD binary pulsars,} in which radio emission is triggered when the magnetic axis of a rotating WD intersects its companion’s wind in the binary orbital plane. We use a 36-year timing baseline to infer the orbital period and binary geometry from radio data alone. The model naturally predicts its intermittent emission and double-pulse structure. Crucially, we show that the beat period between the spin and the orbit matches the observed pulse substructure and polarisation signatures, providing strong support for the model. Applying it to the WD pulsar \jlpt{}, it successfully reproduces the emission profile and geometry as well.
\revised[R1.1,R3.1]{Our results suggest analogous emission-site geometries in these related classes of binary system --- a possibility we extend to the broader LPT / WD pulsar population.}
\end{abstract}

\section{Introduction}

White dwarf binary systems are increasingly recognised as hosts of exotic pulsar-like emission, challenging long-standing assumptions that coherent radio pulses are exclusive to neutron star systems\cite{2017NatAs...1E..29B, 2023NatAs...7..931P}. The discovery of long-period transients (LPTs), characterised by their long periodicities (minutes to hours) compared to the bulk of the pulsar population, further deepens the mystery of how and where pulsed coherent radio emission originates \cite{2022Natur.601..526H, 2023Natur.619..487H}. These intriguing systems are characterised by periodic radio pulses with high polarisation (both linear and circular) and a high surface brightness, pointing to a coherent radio emission process. However, the luminosities of the bright radio pulses in \revised[R1.4]{many of} the $\sim$12 systems discovered thus far exceed their spin-down power given typical pulsar-like assumptions, leaving uncertain the energy source. A strong magnetic field has been suggested to overcome this energy problem, suggesting at first the possibility of them being radio magnetars having perhaps had large fall-back at birth slowing down their periods \cite{2022ApJ...934..184R, 2022MNRAS.513L..68G}. 

Two LPTs, ILT\,J1101$+$55 \cite{2025NatAs...9..672D} and GLEAM$-$X\,J0704$-$37 \cite{2024ApJ...976L..21H, 2025A&A...695L...8R}, have recently been associated with WD -- M dwarf (MD) binary systems via the detection of the orbital motion in the optical spectrum of the M companion star, at a period coincident with the radio pulse period. These two systems might be hosting a WD with a spin synchronised with the orbital motion, despite not being in an accretion phase. \revised[R1.2,R1.5]{We note, however, that the radio period need not necessarily equal the WD spin period.} \revised{We compare the aforementioned LPTs to the radio emitting binary WDs (so-called WD pulsars) AR\,Scorpii\cite{2017NatAs...1E..29B} (AR\,Sco) and \jlpt \cite{2023NatAs...7..931P, 2024MNRAS.527.3826P, 2023A&A...674L...9S}}, which are not synchronised with their companions; coherent radio pulses are observed at the WD spin period and/or beat periods resulting from the interaction between the WD spin and the orbital motion, modulated on the orbital period with the brightest pulses arriving at a particular orbital phase. The radio emission in these systems has been intriguing all along, and scenarios involving the dipolar losses of the WD rotating field and the interaction with the MD companion wind have been proposed \cite{2018MNRAS.481.2384P,2022MNRAS.510.2998D,2025ApJ...981...34Q, 2016ApJ...831L..10G,2025arXiv250909224Y}. \revised{A third AR\,Sco-like system, SDSS\,J2306+24\cite{2025MNRAS.tmp.1459C} was also recently discovered with broadly similar characteristics.}

\gpmlpt{} is the longest-lived LPT known, exhibiting a 22-minute period with sporadic pulses observed across 36 years of archival radio data \cite{2023Natur.619..487H}. \revised[R1.8]{The 30 to 300\,s duration broadband radio pulses exhibit both linear and circular polarisation and quasiperiodic substructure.} \revised[R1.4]{Its original classification as a neutron star was based on its coherent radio luminosity ($\sim3.5\times10^{31}$\,erg\,s$^{-1}$ peak isotropic luminosity) but the period derivative $\dot{P} \lesssim 3.6\times10^{-13}$\,s\,s$^{-1}$ \cite{2023Natur.619..487H} puts it in tension with the most generous pulsar emission models \cite{2000ApJ...531L.135Z}. \gpmlpt{} also displays short duration ($\sim$20\,ms) narrow-band sub-pulse features\cite{2025SciA...11P6351M}, suggesting multiple processes.}

Motivated by the detection of a longer periodicity in observations of \gpmlpt{}, we put forward that it is also a WD with a low-mass stellar companion, and propose a geometric model in which radio emission from the WD polar region is modulated by the relative alignment between the WD magnetic moment and the companion star's position. We proceed on the premise that the aforementioned alignment determines the abundance of particles injected from the companion's wind into the WD's magnetosphere \cite{2021NatAs...5..648S, 2024MNRAS.527.3826P}. These particles are accelerated along magnetic field lines, producing a collimated radio beam aligned with the magnetic moment of the WD. The model incorporates orbital inclination, magnetic obliquity, and beam geometry, and predicts observable intensity variations as a function of the spin and orbital phase. Further intensity variations observed in the long-term radio outbursts of LPTs might be ascribed to the intrinsic variability of the companion's wind.

\revised[R1.1,R2.1,R3.1]{Similar models have been proposed for \jlpt{} \cite{2024MNRAS.527.3826P} and AR\,Sco \cite{2018MNRAS.481.2384P,2022MNRAS.510.2998D}, the latter of which reproduces AR\,Sco's more complex pulse morphology with spin and orbit modulated components resembling \gpmlpt{}. The phenomenological similarities between these sources and the geometric similarity invite an analogous physical framework.}

\section{Results}

In a coordinated campaign using MeerKAT, ASKAP, and VLA in 2024, we obtained a near-contiguous 36 hour track of \gpmlpt{}, in which we identified a higher-order periodicity: pulses cluster in paired groups occurring every $\sim$8.75\,hr. We interpret this as an orbital period \gpmlptT{}, which we are able to measure to such low uncertainty by taking advantage of the 36 year baseline (see \autoref{subsec:timing_analysis}).

\begin{figure}[H]
\centering
\includegraphics[width=\linewidth]{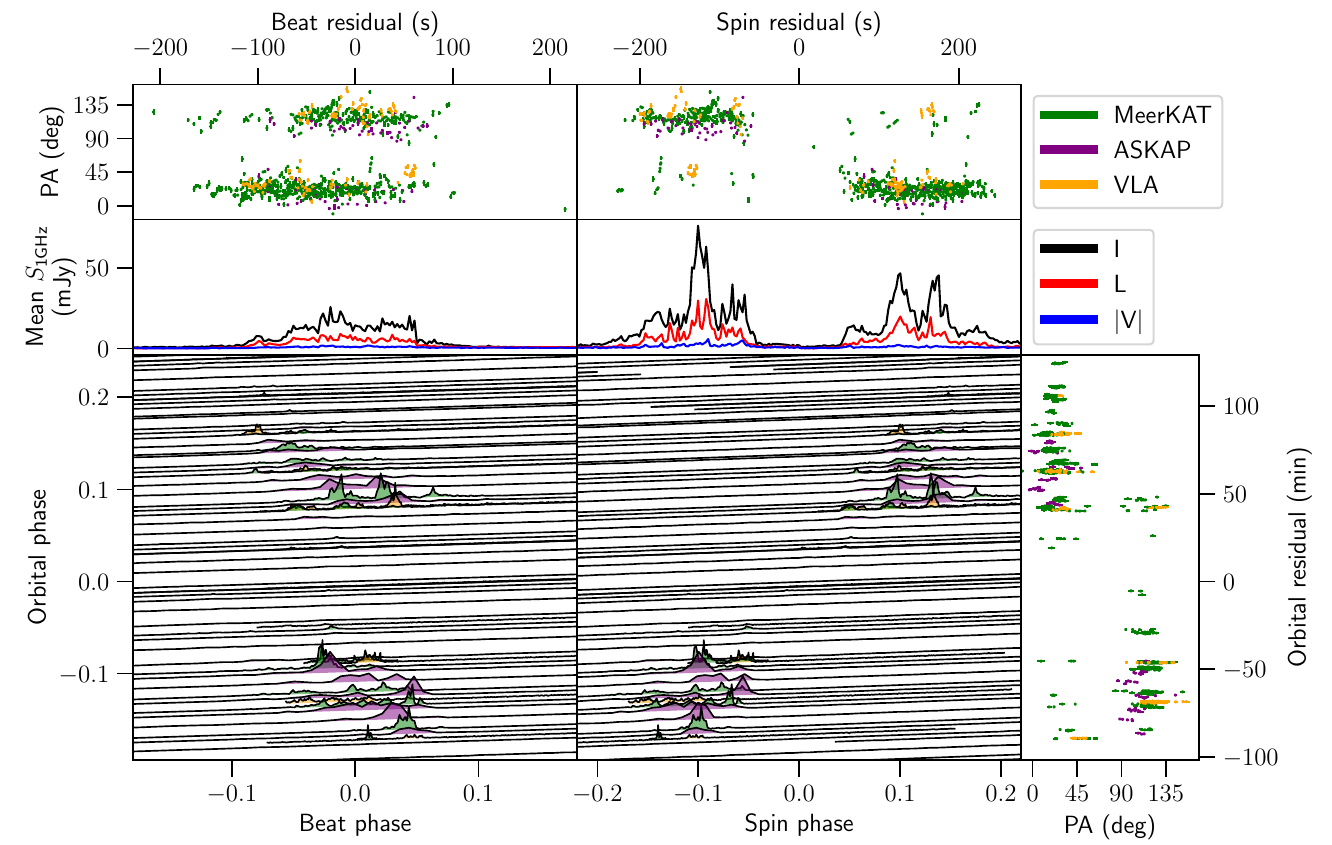}
\caption{Dynamic pulse profile and polarisation position angles of \gpmlpt{}. Profiles are folded vertically on the orbital period \gpmlptT{} and horizontally on \gpmlptPa{} (left) and \gpmlptPb{} (right). Above, the mean flux is calculated over two boxes enclosing the pulse groups: $\lbrace-0.18 < \text{orbital phase} < 0.18\rbrace$ for period A, $(\lbrace-0.18 < \text{orbital phase} < -0.06\rbrace \cap \lbrace\text{spin phase} < 0\rbrace) \cup (\lbrace0.06 < \text{orbital phase} < 0.18\rbrace \cap \lbrace\text{spin phase} > 0\rbrace)$ for period B. \label{fig:period_comparison}}
\end{figure}

Folding the data over this 8.75\,hr modulation period and the previously reported\cite{2023Natur.619..487H} 22-minute period ($P_1$), as well as the beat period ($P_2$) between $P_1$ and $P_\text{orb}$, creates the dynamic pulse profiles in \autoref{fig:period_comparison}, revealing that the pulse substructure aligns between pulses with \gpmlptPb{}, rather than \gpmlptPa{}.
The substructure timescale of $\sim$50 s is consistent throughout the archival data, first discovered by autocorrelation between pulses \cite{2023Natur.619..487H}, and folding on $P_2$ reveals further correlation and vertical alignment of the brightest peaks. This suggests that $P_2$ corresponds to the true spin period of the white dwarf (while $P_1$ is a spin-orbit beat period) inviting its interpretation as the cross-section of the radio beam.
\revised{Unfortunately, none of these periodicities could be confirmed in the optical bands (as for other LPTs) since the high absorption within the Galaxy prevented a firm optical identification. The identification of $P_2$ as the spin period is further corroborated by the agreement of the proposed model with its associated dynamic pulse profile, as opposed to $P_1$, as well as the separation of the orthogonal polarisation modes in spin phase.}

\subsection{Modelling}\label{subsec:model}

\begin{figure}[H]
\centering
\includegraphics[width=\linewidth]{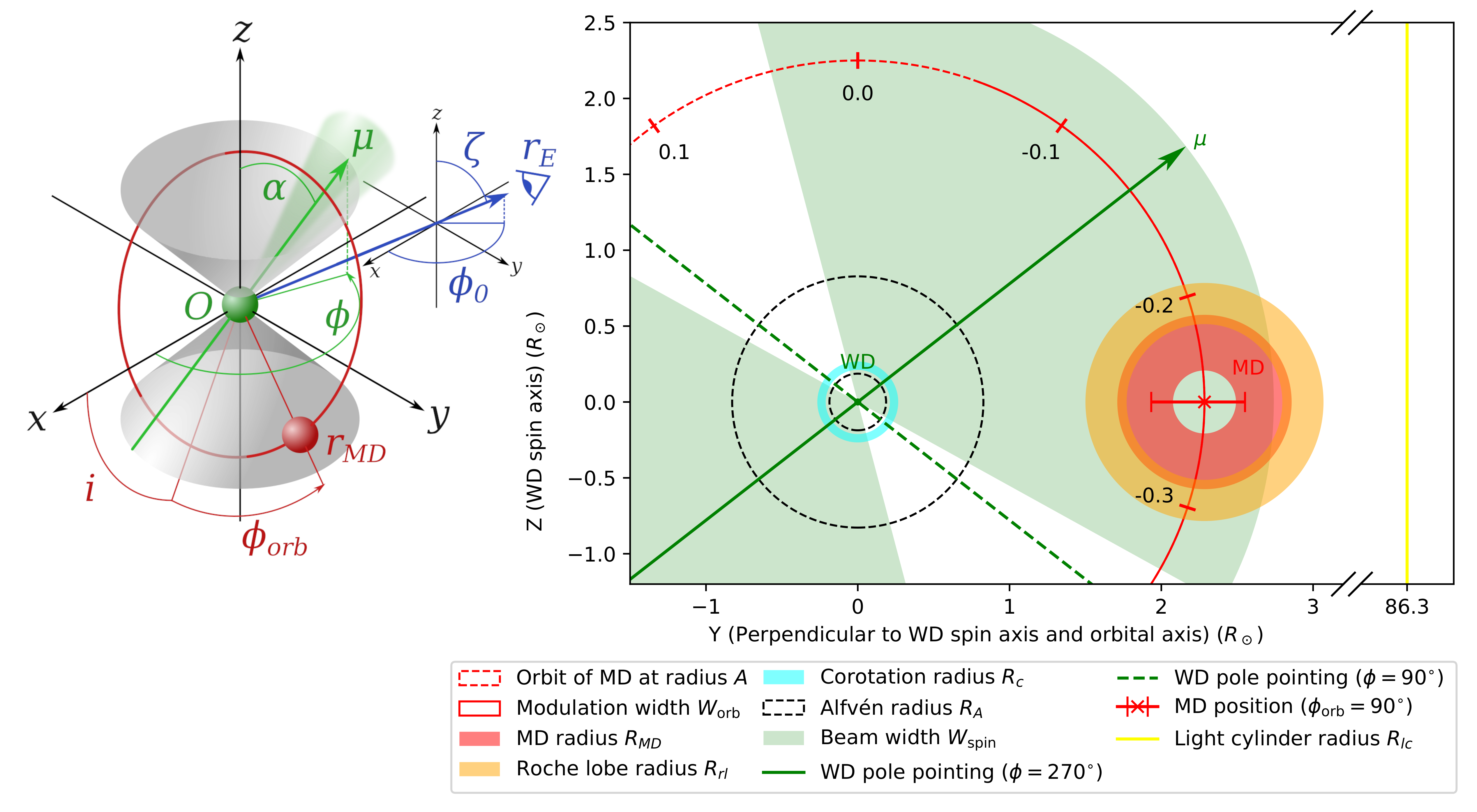}
\caption{Diagram of the binary system, in a moving reference frame centred on the WD.
At left is a not-to-scale diagram of the geometric parameters. In green is WD with its magnetic moment vector. In red, the MD with its orbital path. The blue vector points towards Earth.
At right is a to-scale projection of \gpmlpt{} on the y-z plane, with orbital phases marked. The cone traced by $\hat{\mu}$ crosses the y-z plane at the green full and dashed lines.
The shaded red, orange, and cyan regions cover the ranges of $R_\text{MD}$, $R_{rl}$, and $R_c$ respectively, given the MD mass is $M_\text{MD} \in [0.14, 0.5] M_\odot$ and the WD mass is $M_\text{WD} \in [0.6, 1.2] M_\odot$. The green sector and solid red line are $W_\text{spin}$ and $W_\text{orb}$ respectively. The small and large black dashed circles are the Alfvén radii for the minimum and maximum WD masses respectively, assuming a stellar mass loss rate of $\dot{M}_\text{MD}$ = $10^{-14}$ M$_\odot$ yr$^{-1}$ \cite{2005ApJ...628L.143W}. \label{fig:scale_diagram}}
\end{figure}

The phase difference between the left and right pulse groups in \autoref{fig:period_comparison} cannot be attributed to the difference in light travel time between orbital phases because that would be on the order of seconds for an orbit of this size, not minutes. Instead, we consider a WD with a magnetic moment $\vec{\mu}$ misaligned from its spin axis $\hat{z}$ by angle $\alpha$ and a companion low-mass star (assumed to be an MD for simplicity) in a circular orbit whose axis is inclined by angle $i$ from $\hat{z}$ about $\hat{y}$ (see \autoref{fig:scale_diagram}). When the spin and orbital axes are misaligned, there are two orbital phases at which the rotating magnetic moment vector intersects the MD, which we suggest correspond to the observed pulse groups. Therefore, we model the observed radio emission as modulated by
\begin{enumerate}
    \item a Gaussian function of the angle $\beta_{\text{MD}}$ between the WD magnetic moment and the MD, governing the availability of the MD's wind at the WD pole, and 
    \item a Gaussian function of the angle $\beta$ between the WD magnetic moment and the line of sight (LOS), which is the cross-section of the radio beam parallel to the WD magnetic moment.
\end{enumerate}
This model, although simplified, provides predictive power: emission is expected when both $\beta_{\text{MD}}$ and $\beta$ are small. The resulting phase-space of emission naturally leads to double-peaked pulse groups and intermittent visibility consistent with observations.

\begin{figure}[H]
\centering
\includegraphics[width=\linewidth]{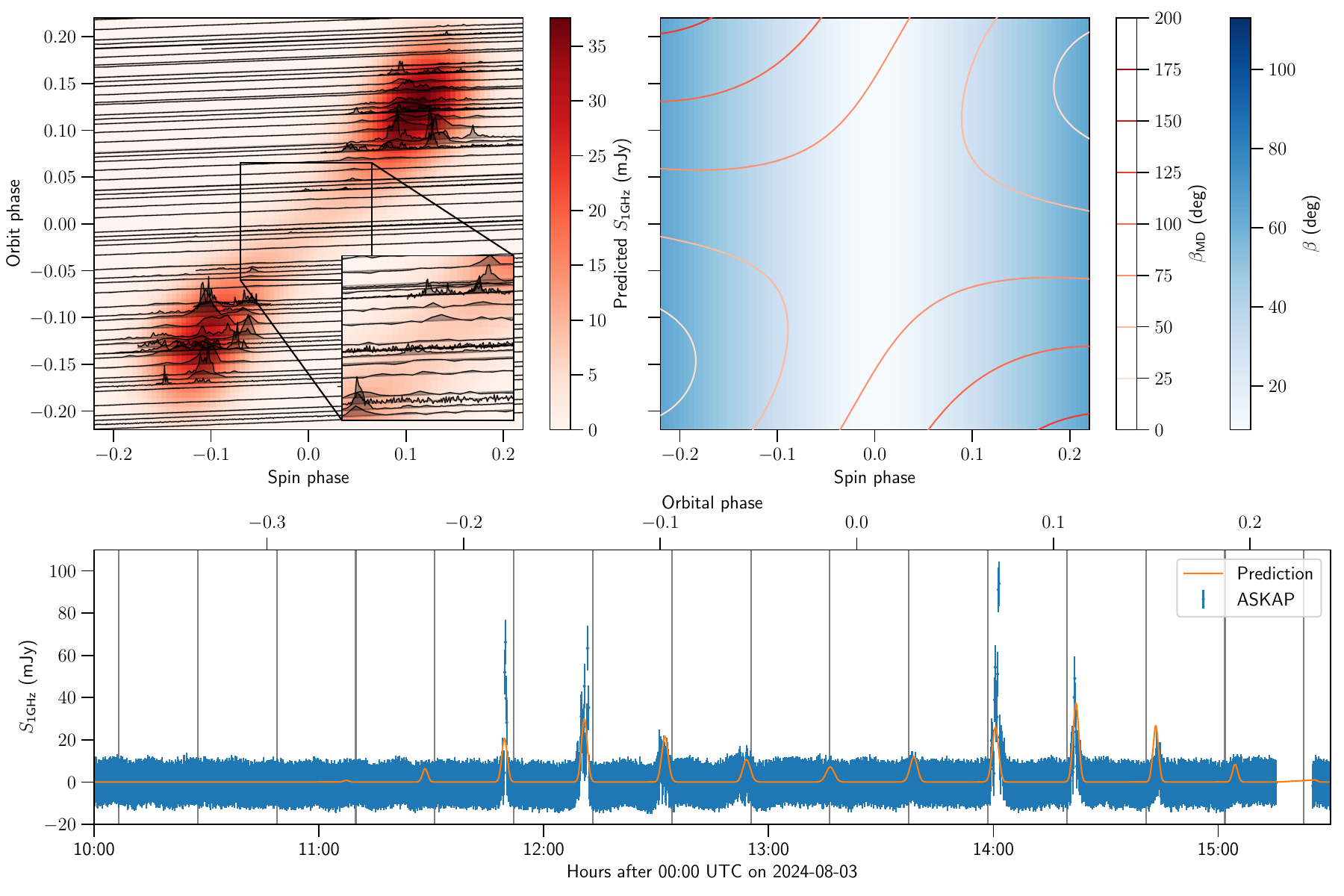}
\caption{Modelled dynamic pulse profile of \gpmlpt{}. At top left, the flux density predicted by the model $I_\text{pred}$ using the best-fit parameters found using MCMC is overlaid on the real pulse profiles. At top right, the colourmap is the LOS-beam angle $\beta$ as a function of spin-orbit phase, and the contours are the beam-MD angle $\beta_\text{MD}$. The bottom panel is a full orbit recorded by ASKAP normalised to 1\,GHz and the associated model prediction. The vertical lines are spaced by the spin period. \label{fig:period_b_predicted_flux}}
\end{figure}

Fitting the model for \gpmlpt{} using a Markov chain Monte Carlo method, we find a near-perpendicular orbital inclination \gpmlpti{} and magnetic obliquity \gpmlptalpha{}. Our LOS has azimuth \gpmlptphiz{} and zenith \gpmlptzeta{}. The radio beam has an opening angle of \gpmlptWspin{} and becomes active when the MD is within \gpmlptWorb{} of $\hat{\mu}$.
\autoref{fig:period_b_predicted_flux} is the predicted dynamic pulse profile, alongside a map of $\beta$ and $\beta_\text{MD}$, and the lightcurve of a representative orbit. Intuitively, the modelled pulse groups are the regions where the $\beta_\text{MD}$ contours cut into the central column of the $\beta$ colour map --- where the Earth, MD, and WD pole are sufficiently aligned. The model also predicts weak emission in the central spin phase $\sim$ orbit phase $\sim$ 0 region, where we indeed see sporadic dim pulses at about the substructure timescale. No good fit exists for $P_1$ as the spin period.
In \autoref{fig:scale_diagram} we show the derived beam angle and orbit of the system together with critical radii on a 2D projection.

\begin{figure}[H]
\centering
\includegraphics[width=\linewidth]{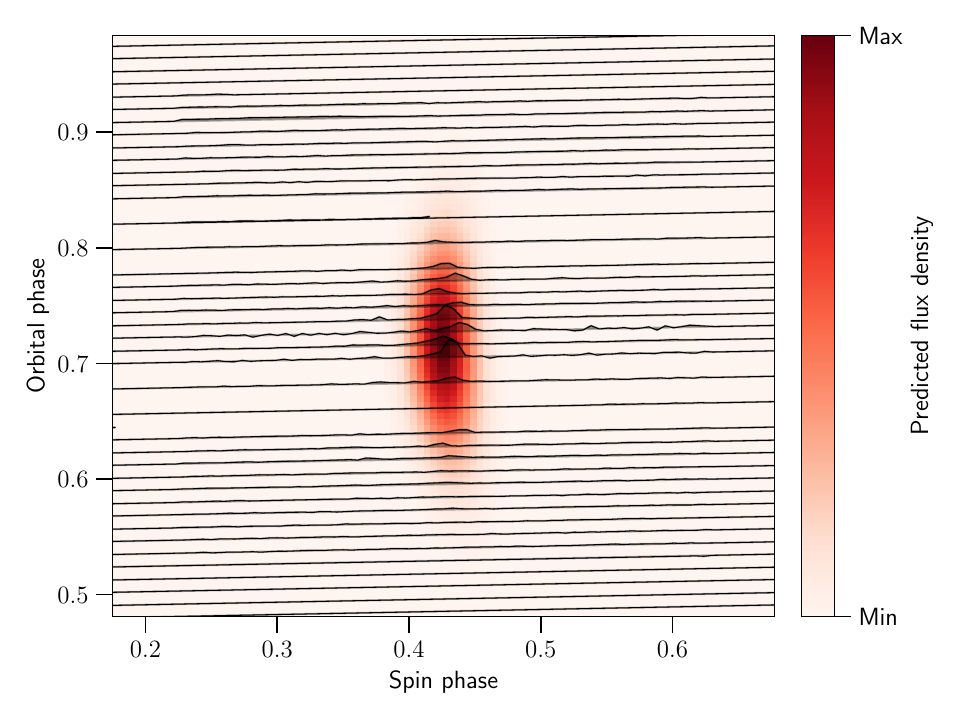}
\caption{Modelled dynamic pulse profile for \jlpt{}.\label{fig:j1912}}
\end{figure}

To test the generality of the model, we applied it to previously published \cite{2023NatAs...7..931P} 7.5 hours of MeerKAT observations of the WD binary pulsar \jlpt{}. Folding the data over the 5.3 minute spin and 4.03 hour orbital periods forms the vertically aligned dynamic pulse profile in \autoref{fig:j1912}. We used a prior of $i = 59 \pm 6^\circ$ based on the inclination angle constraints as measured in the optical band \cite{2024MNRAS.527.3826P}. Given the prior, we find the following constraints on the geometry: \jlpti{}, \jlptalpha{}, \jlptphiz{}, \jlptzeta{}, \jlptWspin{}, and \jlptWorb{}. These results agree broadly with the earlier proposed geometry \cite{2024MNRAS.527.3826P}, but with only one pulse group the geometrical parameters are significantly less constrained than for \gpmlpt{}. The degeneracy of the \jlpt{} parameters suggests that the vertical single-pulse-group morphology should be the most commonly observed among WD pulsars and LPTs.

\section{Discussion}

\revised[R2.2]{Strengthening the case for a WD LPT progenitor beyond the thus-far mentioned are the LPTs ASKAP\,J1448\ensuremath{-}68 \cite{2025MNRAS.542.1208A} and CHIME/ILT\,J1634\ensuremath{+}44 \cite{2025ApJ...988L..29D, 2025A&A...699A.341B}, both of which have been shown to have optical spectral properties consistent with a WD. Furthermore, hints of accretion in the former and pulse arrival time properties in the latter imply possible binary systems.}
\revised{Here we show for \gpmlpt{} that the presence of binarity can be constrained by modelling the radio pulse profiles and arrival times alone, even when a multi-wavelength counterpart is not detected.}

For \gpmlpt{}, \autoref{fig:scale_diagram} shows how the rotating WD magnetosphere is naturally engulfed by the MD wind, possibly being the seed particles that are accelerated by the magnetic field lines generating the pulsed radio emission \cite{2017ApJ...851..143T}. It further shows that: 1) the whole system is inside the WD light cylinder, 2) the companion star, if assumed to be an MD (as for the other two LPTs), is not filling its Roche-Lobe, and 3) depending on the assumed M-dwarf mass loss rate ($\sim10^{-14}$ M$_\odot$ yr$^{-1}$ \cite{2005ApJ...628L.143W}), the WD Alfv\'en radius might be inside the corotation radius, implying that during stellar flares the system may undergo accretion episodes (possibly showing transient X-ray emission as observed in the LPT ASKAP\,J1832\ensuremath{-}0911 \cite{2025Natur.642..583W}, \revised[R1.4,R3.6]{although the X-ray luminosity limit of $L_X < 2\times 10^{33}$\,erg\,s$^{-1}$\cite{2023Natur.619..487H} for \gpmlpt{} does not preclude some level of accretion\cite{2017PASP..129f2001M}), while being in propeller most of the time \cite{2020arXiv200411474L, 2024MNRAS.527.3826P}. ASKAP\,J1448\ensuremath{-}68 has also been observed to emit X-rays\cite{2025MNRAS.542.1208A}, and its multi-wavelength properties are consistent with a magnetic white dwarf, possibly an accreting binary system. In general, it seems that significant accretion may not be a necessary component of the radio emission mechanism (whatever it may be) since we both do \cite{2025Natur.642..583W, 2025MNRAS.542.1208A, 2025ApJ...988L..29D, 2025A&A...699A.341B, 2023NatAs...7..931P} and do not \cite{2025NatAs...9..672D, 2024ApJ...976L..21H, 2017NatAs...1E..29B, 2025MNRAS.tmp.1459C} see signs of such across the LPT / WD pulsar population.}

\revised[1.8,R2.1,3.1]{It has been shown \cite{2018MNRAS.481.2384P,2022MNRAS.510.2998D} that for AR Sco's radio emission, like for that of \gpmlpt{} in this work, the pulse morphology is best modelled with a beam direction locked to the WD rotating frame, as opposed to the binary frame, and modulated by the companion's location in that frame. One may expect the geometric features between the discussed LPTs and WD pulsars to be similar simply because both originate from binary systems, but the comparison is stronger than that. These geometric constraints give us the important clue that the emission direction is dominated by the WD's local field, while powered by interaction with the MD.} If this is not the case it becomes difficult to explain the paired pulse group morphology, and it becomes unclear why the pulse substructure would be related to the spin phase. \revised[R1.9]{This is difficult to reconcile with models like electron cyclotron maser emission (ECME) \cite{1969ApJ...156...59G} or models which place the emission site near the MD \cite{2016ApJ...831L..10G,2017ApJ...835..150K}.}

A scenario which has been suggested for AR\,Sco \cite{2017NatAs...1E..29B, 2017ApJ...851..143T} which allows for an emission region locked to the WD involves relativistic electrons accelerating from the companion towards the WD along the closed magnetic field lines when the companion star's surface is heated by magnetic interaction. The electrons emit synchrotron radiation as they reach magnetic mirror points on their way to the WD pole.
This idea was used successfully in the earlier mentioned geometric model \cite{2018MNRAS.481.2384P} to predict polarimetric properties of AR\,Sco.
\revised[R1.8]{The broadband nature and spectral shape of \gpmlpt{} could be consistent with this, but the radio luminosity is orders of magnitude brighter. Despite the possibility that the radio emission from AR\,Sco and \gpmlpt{} are powered by different mechanisms, we highlight the consistent interpretation in the two systems that the emission is modulated as the source of particles (i.e. the MD) moves along its orbit. If the mechanisms differ then it is likely in the acceleration mechanism of said particles.}

\revised[R1.9]{Alternative forms of ECME in LPTs have been proposed and remain viable, such as relativistic ECME \cite{2025ApJ...981...34Q} or loss-cone-driven maser (LCDM) \cite{2025arXiv250909224Y}, which produce beamed emission aligned with the local magnetic field of the WD.}
\revised[R3.5]{Less luminous ($L \lesssim 10^{25}$\,erg\,s$^{-1}$) radio emission from a related class of WD binaries, (magnetic) cataclysmic variables (CV / mCV), has been observed\cite{2015MNRAS.451.3801C, 2020AdSpR..66.1226B}.
If this emission is ECME, consistent with their typical circular polarisation \cite{1993P&SS...41..333M} (not a requirement in the relativistic case), then there could be a natural spectrum from non-relativistic to relativistic regimes.}

\revised[R1.6,R1.9,R3.2]{In the case of ILT\,J1101\ensuremath{+}55, relativistic ECME was proposed to be powered by unipolar induction, in which an electric field is induced in the MD by the penetrating WD magnetic field, leading to the establishment of a current loop along the field lines connecting the two. However, this requires that the orbital separation is sufficiently small that the WD magnetic field dominates at the location of the MD. According to recent theoretical work \cite{2025arXiv250909224Y}, when the orbit is greater than $P_\text{orb}\sim3$\,hr, the MD magnetic field is strong enough to prevent the WD field from penetrating its surface, and we transition from the unipolar induction phase to the magnetospheric interaction phase. In this phase, radio emission occurs along magnetic field lines connecting the WD to a magnetic reconnection region between the two, rather than the MD itself. The orbit of \gpmlpt{} would place it in the magnetospheric interaction phase, and we calculate that the observed luminosity can be reached without requiring a strongly magnetic WD.}

In any case, the emission mechanism must differ substantially from the common radio pulsar curvature radiation process, a testable prediction of which is the beam opening angle confined by the last open field lines at the pole. Assuming that the emission site is within the orbital radius ($r_{em} \leq A \leq 2.55 R_\odot$) places a generous limit on the half-opening angle of $\rho \leq 15^\circ$, inconsistent with the best-fit half-opening angle $W_\text{spin} = 65 \pm 3 ^\circ$. \revised{Such a wide opening angle is challenging for any model confining the beam between field lines, but the beam morphology is poorly understood at this stage.}

\subsection{Polarisation}

The emission of \gpmlpt{} is consistently highly linearly polarised (up to 100\%) with occasional 90$^\circ$ jumps in polarisation angle (PA) to orthogonal polarisation modes (OPMs). However, \autoref{fig:period_comparison} reveals that the two pulse groups which appear at different orbital and spin phases are dominated by different OPMs, with a similar amount of variation in each group ($\sigma_\text{PA} \approx {\sim}10^\circ$). The PAs of individual pulses evolve more gradually and systematically across spin phase than can be seen in the point clusters in \autoref{fig:period_comparison} (see \cite{2025SciA...11P6351M}), but we focus on the average behaviour.
Both the single pulse PA variations and the occasional OPM jumps within a pulse group may be due to turbulence in the intrabinary wind, \revised[R1.10]{but why one OPM should be preferred at a particular spin/orbit phase is unclear. Other spectral characteristics (dispersion measure (DM), rotation measure (RM), spectral shape) remain constant across the orbit, although DM and RM variations can be suppressed by the surrounding magnetospheric plasma.}

It can further be seen that there is a small downturn in the PAs at around spin phase $-0.05$ (the trailing edge of the first group), as well as a similar upturn in the PAs at spin phase $0.05$ (the leading edge of the second group).
Despite the possibility that dim pulses arriving between the two groups could potentially connect the PAs smoothly across intermediate spin--orbit phases (with greater sensitivity), the flatness of the PAs of the two groups themselves, as well as the occasional $90^\circ$ jump within each group, lead us to prefer an OPM interpretation over a rotating vector model\cite{1969ApL.....3..225R}.

Although not well understood, OPMs in pulsars are thought to be magnetospheric in origin, possibly due to birefringent properties of the co-rotating plasma\cite{1997A&A...327..155G}. We advance a similarly broad interpretation here: \revised[R1.10]{the linear polarisation is dominated by the average magnetic properties of the plasma through which the WD beam propagates along the LOS, switching the visible polarisation mode with spin phase.
If the mechanism behind the pulse group PA separation and the short duration OPMs are related, then further study of \gpmlpt{} may give insights into OPMs more broadly.}

\section{Conclusion}

Future work will include applying the model to the growing number of LPTs, which will not only test it, but serve to constrain the distribution of their geometries. Of particular interest is the $P=$6.45-h LPT ASKAP\,J183950.5\ensuremath{-}075635.0 \cite{2025NatAs...9..393L} which has timing variation and pulse profile evolution phenomenologically similar to \gpmlpt{} but with an interpulse, extending the geometric parameter space. Also of interest would be ASKAP\,J1755\ensuremath{-}2527, recently confirmed to be a 1.16-hour intermittent LPT \cite{2025MNRAS.542..203M}, the emission-state-switching 54 minute LPT ASKAP\,J1935+2148 \cite{2024NatAs...8.1159C} and CHIME\,J0630\ensuremath{+}25 \cite{2025ApJ...990L..49D}, whose 421 s period makes it an interesting counterpart to \jlpt{}.

Our results demonstrate that LPTs like \gpmlpt{} can be understood within the same emission geometry framework as radio emitting WD binary pulsars such as \jlpt{} and AR\,Sco.
The diversity in the observed periods, pulse morphologies, and polarisation states arise from variations in orbital synchronisation and viewing geometry.
\revised{Of course, conclusions made by drawing from the properties of the whole disparate population of LPTs rely on the assumption that they are the same kind of system, which remains uncertain; many LPTs are still consistent with a magnetar \cite{2022Natur.601..526H, 2023JApA...44....1K, 2023ApJ...943....3T} (or other \cite{2022Ap&SS.367..108K,2025arXiv250710682C,2025ApJ...988L..11M,2024PhRvD.109f3004B,2024ApJ...972...60X,2025arXiv250617389N,2025ApJ...986...98Z}) interpretation.
We have shown that careful examination of radio timing properties can help disambiguate the nature of LPTs.}

The possible evolutionary connection between these systems remains debated, and we cannot exclude them being at different evolutionary stages but with common properties and geometries allowing them to emit bright polarised radio emission. With enough samples, we expect their evolutionary stages to emerge, and the methods developed in this work will enable insight into binary WD and magnetic field evolution in a way difficult to select from optical surveys alone.


\bibliography{refs.bib}

\begin{thebibliography}{10}
\expandafter\ifx\csname url\endcsname\relax
  \def\url#1{\texttt{#1}}\fi
\expandafter\ifx\csname urlprefix\endcsname\relax\def\urlprefix{URL }\fi
\providecommand{\bibinfo}[2]{#2}
\providecommand{\eprint}[2][]{\url{#2}}

\bibitem{2017NatAs...1E..29B}
\bibinfo{author}{{Buckley}, D.~A.~H.}, \bibinfo{author}{{Meintjes}, P.~J.}, \bibinfo{author}{{Potter}, S.~B.}, \bibinfo{author}{{Marsh}, T.~R.} \& \bibinfo{author}{{G{\"a}nsicke}, B.~T.}
\newblock \bibinfo{title}{{Polarimetric evidence of a white dwarf pulsar in the binary system AR Scorpii}}.
\newblock \emph{\bibinfo{journal}{Nature Astronomy}} \textbf{\bibinfo{volume}{1}}, \bibinfo{pages}{0029} (\bibinfo{year}{2017}).

\bibitem{2023NatAs...7..931P}
\bibinfo{author}{{Pelisoli}, I.} \emph{et~al.}
\newblock \bibinfo{title}{{A 5.3-min-period pulsing white dwarf in a binary detected from radio to X-rays}}.
\newblock \emph{\bibinfo{journal}{Nature Astronomy}} \textbf{\bibinfo{volume}{7}}, \bibinfo{pages}{931--942} (\bibinfo{year}{2023}).

\bibitem{2022Natur.601..526H}
\bibinfo{author}{{Hurley-Walker}, N.} \emph{et~al.}
\newblock \bibinfo{title}{{A radio transient with unusually slow periodic emission}}.
\newblock \emph{\bibinfo{journal}{Nature}} \textbf{\bibinfo{volume}{601}}, \bibinfo{pages}{526--530} (\bibinfo{year}{2022}).

\bibitem{2023Natur.619..487H}
\bibinfo{author}{{Hurley-Walker}, N.} \emph{et~al.}
\newblock \bibinfo{title}{{A long-period radio transient active for three decades}}.
\newblock \emph{\bibinfo{journal}{\nat}} \textbf{\bibinfo{volume}{619}}, \bibinfo{pages}{487--490} (\bibinfo{year}{2023}).

\bibitem{2022ApJ...934..184R}
\bibinfo{author}{{Ronchi}, M.}, \bibinfo{author}{{Rea}, N.}, \bibinfo{author}{{Graber}, V.} \& \bibinfo{author}{{Hurley-Walker}, N.}
\newblock \bibinfo{title}{{Long-period Pulsars as Possible Outcomes of Supernova Fallback Accretion}}.
\newblock \emph{\bibinfo{journal}{Astrophys. J.}} \textbf{\bibinfo{volume}{934}}, \bibinfo{pages}{184} (\bibinfo{year}{2022}).

\bibitem{2022MNRAS.513L..68G}
\bibinfo{author}{{Gen{\c{c}}ali}, A.~A.}, \bibinfo{author}{{Ertan}, {\"U}.} \& \bibinfo{author}{{Alpar}, M.~A.}
\newblock \bibinfo{title}{{Evolution of the long-period pulsar GLEAM-X J162759.5-523504.3}}.
\newblock \emph{\bibinfo{journal}{\mnras}} \textbf{\bibinfo{volume}{513}}, \bibinfo{pages}{L68--L71} (\bibinfo{year}{2022}).

\bibitem{2025NatAs...9..672D}
\bibinfo{author}{{de Ruiter}, I.} \emph{et~al.}
\newblock \bibinfo{title}{{Sporadic radio pulses from a white dwarf binary at the orbital period}}.
\newblock \emph{\bibinfo{journal}{Nature Astronomy}} \textbf{\bibinfo{volume}{9}}, \bibinfo{pages}{672--684} (\bibinfo{year}{2025}).

\bibitem{2024ApJ...976L..21H}
\bibinfo{author}{{Hurley-Walker}, N.} \emph{et~al.}
\newblock \bibinfo{title}{{A 2.9 hr Periodic Radio Transient with an Optical Counterpart}}.
\newblock \emph{\bibinfo{journal}{\apjl}} \textbf{\bibinfo{volume}{976}}, \bibinfo{pages}{L21} (\bibinfo{year}{2024}).

\bibitem{2025A&A...695L...8R}
\bibinfo{author}{{Rodriguez}, A.~C.}
\newblock \bibinfo{title}{{Spectroscopic detection of a 2.9-hour orbit in a long-period radio transient}}.
\newblock \emph{\bibinfo{journal}{\aap}} \textbf{\bibinfo{volume}{695}}, \bibinfo{pages}{L8} (\bibinfo{year}{2025}).

\bibitem{2024MNRAS.527.3826P}
\bibinfo{author}{{Pelisoli}, I.} \emph{et~al.}
\newblock \bibinfo{title}{{Unveiling the white dwarf in J191213.72 - 441045.1 through ultraviolet observations}}.
\newblock \emph{\bibinfo{journal}{\mnras}} \textbf{\bibinfo{volume}{527}}, \bibinfo{pages}{3826--3836} (\bibinfo{year}{2024}).

\bibitem{2023A&A...674L...9S}
\bibinfo{author}{{Schwope}, A.} \emph{et~al.}
\newblock \bibinfo{title}{{X-ray properties of the white dwarf pulsar eRASSU J191213.9{\ensuremath{-}}441044}}.
\newblock \emph{\bibinfo{journal}{\aap}} \textbf{\bibinfo{volume}{674}}, \bibinfo{pages}{L9} (\bibinfo{year}{2023}).

\bibitem{2018MNRAS.481.2384P}
\bibinfo{author}{{Potter}, S.~B.} \& \bibinfo{author}{{Buckley}, D. A.~H.}
\newblock \bibinfo{title}{{Time series photopolarimetry and modelling of the white dwarf pulsar in AR Scorpii}}.
\newblock \emph{\bibinfo{journal}{\mnras}} \textbf{\bibinfo{volume}{481}}, \bibinfo{pages}{2384--2392} (\bibinfo{year}{2018}).

\bibitem{2022MNRAS.510.2998D}
\bibinfo{author}{{du Plessis}, L.} \emph{et~al.}
\newblock \bibinfo{title}{{Probing the non-thermal emission geometry of AR Sco via optical phase-resolved polarimetry}}.
\newblock \emph{\bibinfo{journal}{\mnras}} \textbf{\bibinfo{volume}{510}}, \bibinfo{pages}{2998--3010} (\bibinfo{year}{2022}).

\bibitem{2025ApJ...981...34Q}
\bibinfo{author}{{Qu}, Y.} \& \bibinfo{author}{{Zhang}, B.}
\newblock \bibinfo{title}{{Magnetic Interactions in White Dwarf Binaries as Mechanism for Long-period Radio Transients}}.
\newblock \emph{\bibinfo{journal}{\apj}} \textbf{\bibinfo{volume}{981}}, \bibinfo{pages}{34} (\bibinfo{year}{2025}).

\bibitem{2016ApJ...831L..10G}
\bibinfo{author}{{Geng}, J.-J.}, \bibinfo{author}{{Zhang}, B.} \& \bibinfo{author}{{Huang}, Y.-F.}
\newblock \bibinfo{title}{{A Model of White Dwarf Pulsar AR Scorpii}}.
\newblock \emph{\bibinfo{journal}{\apjl}} \textbf{\bibinfo{volume}{831}}, \bibinfo{pages}{L10} (\bibinfo{year}{2016}).

\bibitem{2025arXiv250909224Y}
\bibinfo{author}{{Yang}, Y.-P.}
\newblock \bibinfo{title}{{Magnetic White Dwarf -- M Dwarf Binaries in Pre-polar Phase as Special Population of Long-Period Radio Transients}}.
\newblock \emph{\bibinfo{journal}{arXiv e-prints}} \bibinfo{pages}{arXiv:2509.09224} (\bibinfo{year}{2025}).

\bibitem{2025MNRAS.tmp.1459C}
\bibinfo{author}{{Castro Segura}, N.} \emph{et~al.}
\newblock \bibinfo{title}{{A Sibling of AR Scorpii: SDSS J230641.47+244055.8 and the Observational Blueprint of White Dwarf Pulsars}}.
\newblock \emph{\bibinfo{journal}{\mnras}}  (\bibinfo{year}{2025}).

\bibitem{2000ApJ...531L.135Z}
\bibinfo{author}{{Zhang}, B.}, \bibinfo{author}{{Harding}, A.~K.} \& \bibinfo{author}{{Muslimov}, A.~G.}
\newblock \bibinfo{title}{{Radio Pulsar Death Line Revisited: Is PSR J2144-3933 Anomalous?}}
\newblock \emph{\bibinfo{journal}{Astrophys. J. Lett.}} \textbf{\bibinfo{volume}{531}}, \bibinfo{pages}{L135--L138} (\bibinfo{year}{2000}).

\bibitem{2025SciA...11P6351M}
\bibinfo{author}{{Men}, Y.}, \bibinfo{author}{{McSweeney}, S.}, \bibinfo{author}{{Hurley-Walker}, N.}, \bibinfo{author}{{Barr}, E.} \& \bibinfo{author}{{Stappers}, B.}
\newblock \bibinfo{title}{{A highly magnetized long-period radio transient exhibiting unusual emission features}}.
\newblock \emph{\bibinfo{journal}{Science Advances}} \textbf{\bibinfo{volume}{11}}, \bibinfo{pages}{eadp6351} (\bibinfo{year}{2025}).

\bibitem{2021NatAs...5..648S}
\bibinfo{author}{{Schreiber}, M.~R.}, \bibinfo{author}{{Belloni}, D.}, \bibinfo{author}{{G{\"a}nsicke}, B.~T.}, \bibinfo{author}{{Parsons}, S.~G.} \& \bibinfo{author}{{Zorotovic}, M.}
\newblock \bibinfo{title}{{The origin and evolution of magnetic white dwarfs in close binary stars}}.
\newblock \emph{\bibinfo{journal}{Nature Astronomy}} \textbf{\bibinfo{volume}{5}}, \bibinfo{pages}{648--654} (\bibinfo{year}{2021}).

\bibitem{2005ApJ...628L.143W}
\bibinfo{author}{{Wood}, B.~E.}, \bibinfo{author}{{M{\"u}ller}, H.~R.}, \bibinfo{author}{{Zank}, G.~P.}, \bibinfo{author}{{Linsky}, J.~L.} \& \bibinfo{author}{{Redfield}, S.}
\newblock \bibinfo{title}{{New Mass-Loss Measurements from Astrospheric Ly{\ensuremath{\alpha}} Absorption}}.
\newblock \emph{\bibinfo{journal}{\apjl}} \textbf{\bibinfo{volume}{628}}, \bibinfo{pages}{L143--L146} (\bibinfo{year}{2005}).

\bibitem{2025MNRAS.542.1208A}
\bibinfo{author}{{Anumarlapudi}, A.} \emph{et~al.}
\newblock \bibinfo{title}{{ASKAP J144834-685644: a newly discovered long period radio transient detected from radio to X-rays}}.
\newblock \emph{\bibinfo{journal}{\mnras}} \textbf{\bibinfo{volume}{542}}, \bibinfo{pages}{1208--1232} (\bibinfo{year}{2025}).

\bibitem{2025ApJ...988L..29D}
\bibinfo{author}{{Dong}, F.~A.} \emph{et~al.}
\newblock \bibinfo{title}{{CHIME/Fast Radio Burst Discovery of an Unusual Circularly Polarized Long-period Radio Transient with an Accelerating Spin Period}}.
\newblock \emph{\bibinfo{journal}{\apjl}} \textbf{\bibinfo{volume}{988}}, \bibinfo{pages}{L29} (\bibinfo{year}{2025}).

\bibitem{2025A&A...699A.341B}
\bibinfo{author}{{Bloot}, S.} \emph{et~al.}
\newblock \bibinfo{title}{{Strongly polarised radio pulses from a new white-dwarf-hosting long-period transient}}.
\newblock \emph{\bibinfo{journal}{\aap}} \textbf{\bibinfo{volume}{699}}, \bibinfo{pages}{A341} (\bibinfo{year}{2025}).

\bibitem{2017ApJ...851..143T}
\bibinfo{author}{{Takata}, J.}, \bibinfo{author}{{Yang}, H.} \& \bibinfo{author}{{Cheng}, K.~S.}
\newblock \bibinfo{title}{{A Model for AR Scorpii: Emission from Relativistic Electrons Trapped by Closed Magnetic Field Lines of Magnetic White Dwarfs}}.
\newblock \emph{\bibinfo{journal}{\apj}} \textbf{\bibinfo{volume}{851}}, \bibinfo{pages}{143} (\bibinfo{year}{2017}).

\bibitem{2025Natur.642..583W}
\bibinfo{author}{{Wang}, Z.} \emph{et~al.}
\newblock \bibinfo{title}{{Detection of X-ray emission from a bright long-period radio transient}}.
\newblock \emph{\bibinfo{journal}{\nat}} \textbf{\bibinfo{volume}{642}}, \bibinfo{pages}{583--586} (\bibinfo{year}{2025}).

\bibitem{2017PASP..129f2001M}
\bibinfo{author}{{Mukai}, K.}
\newblock \bibinfo{title}{{X-Ray Emissions from Accreting White Dwarfs: A Review}}.
\newblock \emph{\bibinfo{journal}{\pasp}} \textbf{\bibinfo{volume}{129}}, \bibinfo{pages}{062001} (\bibinfo{year}{2017}).

\bibitem{2020arXiv200411474L}
\bibinfo{author}{{Lyutikov}, M.} \emph{et~al.}
\newblock \bibinfo{title}{{Magnetospheric interaction in white dwarf binaries AR Sco and AE Aqr}}.
\newblock \emph{\bibinfo{journal}{arXiv e-prints}} \bibinfo{pages}{arXiv:2004.11474} (\bibinfo{year}{2020}).

\bibitem{1969ApJ...156...59G}
\bibinfo{author}{{Goldreich}, P.} \& \bibinfo{author}{{Lynden-Bell}, D.}
\newblock \bibinfo{title}{{Io, a jovian unipolar inductor}}.
\newblock \emph{\bibinfo{journal}{\apj}} \textbf{\bibinfo{volume}{156}}, \bibinfo{pages}{59--78} (\bibinfo{year}{1969}).

\bibitem{2017ApJ...835..150K}
\bibinfo{author}{{Katz}, J.~I.}
\newblock \bibinfo{title}{{AR Sco: A Precessing White Dwarf Synchronar?}}
\newblock \emph{\bibinfo{journal}{\apj}} \textbf{\bibinfo{volume}{835}}, \bibinfo{pages}{150} (\bibinfo{year}{2017}).

\bibitem{2015MNRAS.451.3801C}
\bibinfo{author}{{Coppejans}, D.~L.} \emph{et~al.}
\newblock \bibinfo{title}{{Novalike cataclysmic variables are significant radio emitters}}.
\newblock \emph{\bibinfo{journal}{\mnras}} \textbf{\bibinfo{volume}{451}}, \bibinfo{pages}{3801--3813} (\bibinfo{year}{2015}).

\bibitem{2020AdSpR..66.1226B}
\bibinfo{author}{{Barrett}, P.}, \bibinfo{author}{{Dieck}, C.}, \bibinfo{author}{{Beasley}, A.~J.}, \bibinfo{author}{{Mason}, P.~A.} \& \bibinfo{author}{{Singh}, K.~P.}
\newblock \bibinfo{title}{{Radio observations of magnetic cataclysmic variables}}.
\newblock \emph{\bibinfo{journal}{Advances in Space Research}} \textbf{\bibinfo{volume}{66}}, \bibinfo{pages}{1226--1234} (\bibinfo{year}{2020}).

\bibitem{1993P&SS...41..333M}
\bibinfo{author}{{Melrose}, D.~B.} \& \bibinfo{author}{{Dulk}, G.~A.}
\newblock \bibinfo{title}{{Electron cyclotron maser emission at oblique angles}}.
\newblock \emph{\bibinfo{journal}{\planss}} \textbf{\bibinfo{volume}{41}}, \bibinfo{pages}{333--339} (\bibinfo{year}{1993}).

\bibitem{1969ApL.....3..225R}
\bibinfo{author}{{Radhakrishnan}, V.} \& \bibinfo{author}{{Cooke}, D.~J.}
\newblock \bibinfo{title}{{Magnetic Poles and the Polarization Structure of Pulsar Radiation}}.
\newblock \emph{\bibinfo{journal}{Astrophys. J. Lett.}} \textbf{\bibinfo{volume}{3}}, \bibinfo{pages}{225} (\bibinfo{year}{1969}).

\bibitem{1997A&A...327..155G}
\bibinfo{author}{{Gangadhara}, R.~T.}
\newblock \bibinfo{title}{{Orthogonal polarization mode phenomenon in pulsars.}}
\newblock \emph{\bibinfo{journal}{\aap}} \textbf{\bibinfo{volume}{327}}, \bibinfo{pages}{155--166} (\bibinfo{year}{1997}).

\bibitem{2025NatAs...9..393L}
\bibinfo{author}{{Lee}, Y.~W.~J.} \emph{et~al.}
\newblock \bibinfo{title}{{The emission of interpulses by a 6.45-h-period coherent radio transient}}.
\newblock \emph{\bibinfo{journal}{Nature Astronomy}} \textbf{\bibinfo{volume}{9}}, \bibinfo{pages}{393--405} (\bibinfo{year}{2025}).

\bibitem{2025MNRAS.542..203M}
\bibinfo{author}{{McSweeney}, S.~J.} \emph{et~al.}
\newblock \bibinfo{title}{{A new long-period radio transient: discovery of pulses repeating every 1.16 h from ASKAP J175534.9‑252749.1}}.
\newblock \emph{\bibinfo{journal}{\mnras}} \textbf{\bibinfo{volume}{542}}, \bibinfo{pages}{203--214} (\bibinfo{year}{2025}).

\bibitem{2024NatAs...8.1159C}
\bibinfo{author}{{Caleb}, M.} \emph{et~al.}
\newblock \bibinfo{title}{{An emission-state-switching radio transient with a 54-minute period}}.
\newblock \emph{\bibinfo{journal}{Nature Astronomy}} \textbf{\bibinfo{volume}{8}}, \bibinfo{pages}{1159--1168} (\bibinfo{year}{2024}).

\bibitem{2025ApJ...990L..49D}
\bibinfo{author}{{Dong}, F.~A.} \emph{et~al.}
\newblock \bibinfo{title}{{CHIME/Fast Radio Burst/Pulsar Discovery of a Nearby Long-period Radio Transient with a Timing Glitch}}.
\newblock \emph{\bibinfo{journal}{\apjl}} \textbf{\bibinfo{volume}{990}}, \bibinfo{pages}{L49} (\bibinfo{year}{2025}).

\bibitem{2023JApA...44....1K}
\bibinfo{author}{{Konar}, S.}
\newblock \bibinfo{title}{{Enigma of GLEAM-X J162759.5{\textendash}523504.3}}.
\newblock \emph{\bibinfo{journal}{Journal of Astrophysics and Astronomy}} \textbf{\bibinfo{volume}{44}}, \bibinfo{pages}{1} (\bibinfo{year}{2023}).

\bibitem{2023ApJ...943....3T}
\bibinfo{author}{{Tong}, H.}
\newblock \bibinfo{title}{{Discussions on the Nature of GLEAM-X J162759.5-523504.3}}.
\newblock \emph{\bibinfo{journal}{\apj}} \textbf{\bibinfo{volume}{943}}, \bibinfo{pages}{3} (\bibinfo{year}{2023}).

\bibitem{2022Ap&SS.367..108K}
\bibinfo{author}{{Katz}, J.~I.}
\newblock \bibinfo{title}{{GLEAM-X J162759.5{\ensuremath{-}}523504.3 as a white dwarf pulsar}}.
\newblock \emph{\bibinfo{journal}{Astrophys. Space Sci.}} \textbf{\bibinfo{volume}{367}}, \bibinfo{pages}{108} (\bibinfo{year}{2022}).

\bibitem{2025arXiv250710682C}
\bibinfo{author}{{Cary}, S.}, \bibinfo{author}{{Lu}, W.}, \bibinfo{author}{{Leung}, C.} \& \bibinfo{author}{{Wong}, T. L.~S.}
\newblock \bibinfo{title}{{Accretion from a Shock-Inflated Companion: Spinning Down Neutron Stars to Hour-Long Periods}}.
\newblock \emph{\bibinfo{journal}{arXiv e-prints}} \bibinfo{pages}{arXiv:2507.10682} (\bibinfo{year}{2025}).

\bibitem{2025ApJ...988L..11M}
\bibinfo{author}{{Mao}, Y.-H.}, \bibinfo{author}{{Li}, X.-D.}, \bibinfo{author}{{Lai}, D.}, \bibinfo{author}{{Deng}, Z.-L.} \& \bibinfo{author}{{Yang}, H.-R.}
\newblock \bibinfo{title}{{A Binary Origin for Ultralong-period Radio Pulsars}}.
\newblock \emph{\bibinfo{journal}{\apjl}} \textbf{\bibinfo{volume}{988}}, \bibinfo{pages}{L11} (\bibinfo{year}{2025}).

\bibitem{2024PhRvD.109f3004B}
\bibinfo{author}{{Baumgarte}, T.~W.} \& \bibinfo{author}{{Shapiro}, S.~L.}
\newblock \bibinfo{title}{{Could long-period transients be powered by primordial black hole capture?}}
\newblock \emph{\bibinfo{journal}{\prd}} \textbf{\bibinfo{volume}{109}}, \bibinfo{pages}{063004} (\bibinfo{year}{2024}).

\bibitem{2024ApJ...972...60X}
\bibinfo{author}{{Xiao}, X.} \& \bibinfo{author}{{Shen}, R.-F.}
\newblock \bibinfo{title}{{Apparently Ultralong Period Radio Signals from Self-lensed Pulsar{\textendash}Black Hole Binaries}}.
\newblock \emph{\bibinfo{journal}{\apj}} \textbf{\bibinfo{volume}{972}}, \bibinfo{pages}{60} (\bibinfo{year}{2024}).

\bibitem{2025arXiv250617389N}
\bibinfo{author}{{Nathanail}, A.}
\newblock \bibinfo{title}{{Identifying Long Radio Transients with Accompanying X-Ray Emission as Disk-Jet Precessing Black Holes: The Case of ASKAP J1832-0911}}.
\newblock \emph{\bibinfo{journal}{arXiv e-prints}} \bibinfo{pages}{arXiv:2506.17389} (\bibinfo{year}{2025}).

\bibitem{2025ApJ...986...98Z}
\bibinfo{author}{{Zhou}, X.}, \bibinfo{author}{{Kurban}, A.}, \bibinfo{author}{{Liu}, W.-T.}, \bibinfo{author}{{Wang}, N.} \& \bibinfo{author}{{Yuan}, Y.-J.}
\newblock \bibinfo{title}{{Nature of Ultralong Period Radio Transients: Could They Be Strange Dwarf Pulsars?}}
\newblock \emph{\bibinfo{journal}{\apj}} \textbf{\bibinfo{volume}{986}}, \bibinfo{pages}{98} (\bibinfo{year}{2025}).

\bibitem{2021PASA...38....9H}
\bibinfo{author}{{Hotan}, A.~W.} \emph{et~al.}
\newblock \bibinfo{title}{{Australian square kilometre array pathfinder: I. system description}}.
\newblock \emph{\bibinfo{journal}{Publ. Astron. Soc. Aust.}} \textbf{\bibinfo{volume}{38}}, \bibinfo{pages}{e009} (\bibinfo{year}{2021}).

\bibitem{2016mks..confE...1J}
\bibinfo{author}{{Jonas}, J.} \& \bibinfo{author}{{MeerKAT Team}}.
\newblock \bibinfo{title}{{The MeerKAT Radio Telescope}}.
\newblock In \emph{\bibinfo{booktitle}{MeerKAT Science: On the Pathway to the SKA}}, \bibinfo{pages}{1} (\bibinfo{year}{2016}).

\bibitem{2011ApJ...739L...1P}
\bibinfo{author}{{Perley}, R.~A.}, \bibinfo{author}{{Chandler}, C.~J.}, \bibinfo{author}{{Butler}, B.~J.} \& \bibinfo{author}{{Wrobel}, J.~M.}
\newblock \bibinfo{title}{{The Expanded Very Large Array: A New Telescope for New Science}}.
\newblock \emph{\bibinfo{journal}{\apjl}} \textbf{\bibinfo{volume}{739}}, \bibinfo{pages}{L1} (\bibinfo{year}{2011}).

\bibitem{2017ApJS..230....7P}
\bibinfo{author}{{Perley}, R.~A.} \& \bibinfo{author}{{Butler}, B.~J.}
\newblock \bibinfo{title}{{An Accurate Flux Density Scale from 50 MHz to 50 GHz}}.
\newblock \emph{\bibinfo{journal}{\apjs}} \textbf{\bibinfo{volume}{230}}, \bibinfo{pages}{7} (\bibinfo{year}{2017}).

\bibitem{perley2022evla}
\bibinfo{author}{Perley, R.}, \bibinfo{author}{Greisen, E.} \& \bibinfo{author}{Hugo, B.}
\newblock \bibinfo{title}{{EVLA Memo 219 Enabling MeerKAT Polarimetric Imaging in AIPS}}.
\newblock \bibinfo{type}{Tech. Rep.}, \bibinfo{institution}{National Radio Astronomy Observatory} (\bibinfo{year}{2022}).

\bibitem{2014MNRAS.444..606O}
\bibinfo{author}{{Offringa}, A.~R.} \emph{et~al.}
\newblock \bibinfo{title}{{WSCLEAN: an implementation of a fast, generic wide-field imager for radio astronomy}}.
\newblock \emph{\bibinfo{journal}{Mon. Not. R. Astron. Soc.}} \textbf{\bibinfo{volume}{444}}, \bibinfo{pages}{606--619} (\bibinfo{year}{2014}).

\bibitem{2024zndo..13626183P}
\bibinfo{author}{{Pritchard}, J.}
\newblock \bibinfo{title}{{askap-vast/dstools: v1.0.0}} (\bibinfo{year}{2024}).

\bibitem{2013PASP..125..306F}
\bibinfo{author}{{Foreman-Mackey}, D.}, \bibinfo{author}{{Hogg}, D.~W.}, \bibinfo{author}{{Lang}, D.} \& \bibinfo{author}{{Goodman}, J.}
\newblock \bibinfo{title}{{emcee: The MCMC Hammer}}.
\newblock \emph{\bibinfo{journal}{\pasp}} \textbf{\bibinfo{volume}{125}}, \bibinfo{pages}{306} (\bibinfo{year}{2013}).

\end{thebibliography}




\section*{Methods}
\label{sec:methods}

\subsection{Data}

To obtain a contiguous 36-hour track of \gpmlpt{}, we employed three telescopes: the Australian Square Kilometre Array Pathfinder (ASKAP; \cite{2021PASA...38....9H}), the MeerKAT telescope \cite{2016mks..confE...1J} in South Africa, and the Karl G. Jansky Very Large Array (VLA \cite{2011ApJ...739L...1P}) in the United States of America. To maximise S/N for this steep-spectrum source, we used the lowest-frequency bands available for each telescope. Details are shown in \autoref{tab:obs}.

\begin{table} 
    \footnotesize
    \centering
    \begin{tabular}{rrccccc}
        \hline
        Start      & End        & Telescope & Frequency (MHz) & Observation ID & Proposal/Project ID \\
        \hline
        2024-08-02                                                                    \vspace{-10pt} \\ 
             09:30 &      18:30 & ASKAP     &  800--1087      & SB64328        & AS113               \\
             15:25 &      19:54 & MeerKAT   &  544--1088      & 1722610713     & SCI-20230907-NH-01  \\
                   & 2024-08-03                                                       \vspace{-10pt} \\
             21:24 &      01:31 & MeerKAT   &  544--1088      & 1722632983     & SCI-20230907-NH-01  \\
        2024-08-03                                                                    \vspace{-10pt} \\
             01:10 &      08:55 & VLA       & 1000--2000      & 46270340       & 24A-493             \\
             09:00 &      17:00 & ASKAP     &  800--1087      & SB64345        & AS113               \\
             15:36 &      21:14 & MeerKAT   &  544--1088      & 1722697769     & SCI-20230907-NH-01  \\
                   & 2024-08-04                                                       \vspace{-10pt} \\
             22:12 &      01:33 & MeerKAT   &  544--1088      &  1722722073    & SCI-20230907-NH-01  \\
          \hline
    \end{tabular}
    \caption{Observing details for the campaign tracking \gpmlpt{}, in chronological order. Dates and times are on a UTC scale. ``Start'' refers to the first sample on target and ``end'' refers to the last (i.e. calibration scans are omitted).\label{tab:obs}}
\end{table}

For ASKAP and MeerKAT we used the standard observatory flagging and calibration.
The VLA data were flagged for obvious RFI and  calibrated in AIPS.
The flux density scale was based on the values determined for 3C286 (J1331+3030) \cite{2017ApJS..230....7P}. The EVPA values are based on the modified values for 3C286, as reported in EVLA Memo 219 \cite{perley2022evla}.
We imaged all data using \textsc{WSClean}\cite{2014MNRAS.444..606O}, masking \gpmlpt{} itself, forming a deep model of the sky for each observation. After subtracting this from the visibilities, we phase-rotated to \gpmlpt{}. Using \textsc{DSTools}\cite{2024zndo..13626183P} we averaged the baseline data, using the standard conventions to convert from instrumental to celestial Stokes, producing a dynamic spectrum for each observation.


\jlpt{} was observed at L-band by MeerKAT under project code DDT-20220620-PW-01 on 2022-06-26, and originally published by \cite{2023NatAs...7..931P}. It was tracked for 16 30-min scans at 208.984-kHz/2-s correlator resolution. We used the provided observatory calibration and re-imaged the data using \textsc{WSClean}, masking \jlpt{} to form a deep model of the sky. After subtracting this model, we averaged the baselines to form dynamic spectra of the pulses, then across the frequency axis to produce light curves. (Dispersion is negligible across this band.)

\subsection{Light curve generation}

The flux densities $S_\nu$ of each frequency channel of the dynamic spectra were scaled to the 1 GHz flux density $S_{1\text{GHz}}$ predicted by the spectral fit previously published \cite{2023Natur.619..487H}. The spectral model is of the form
\begin{equation}\label{eqn:spectrum}
    S_\nu = S_{1\text{GHz}} \left( \frac{\nu}{1\text{GHz}} \right)^\alpha \exp q \left( \log \frac{\nu}{1\text{GHz}} \right)^2
\end{equation}
with $\alpha = -3.17\pm0.06$ and $q = -0.56\pm0.03$.
The shape of the spectrum was not found to vary within uncertainty as either a function of orbital or spin phase. The dynamic spectra were then de-dispersed by $\text{DM} = 273.5\pm2.5$ pc cm$^{-3}$ and averaged over frequency to produce the light curves. The stokes $Q$ and $U$ components were corrected for rotation measure by $\text{RM} = 531.83\pm0.14$ rad m$^2$, which was also not found to vary.
Barycentric correction was applied to the time axis using the DE430 ephemeris for the Solar System barycentre, and the MeerKAT PA was corrected for the parallactic angle.

\subsection{Timing analysis}\label{subsec:timing_analysis}

The \gpmlpt{} spin period \gpmlptPa{} was calculated from the previously reported \gpmlptPb{} as the spin-orbit beat period using
\begin{equation}\label{eqn:beat}
    P_2 = \frac{P_\text{orb}}{P_\text{orb}/P_1 + 1}.
\end{equation}

\begin{figure}[H]
\centering
\includegraphics[width=\linewidth]{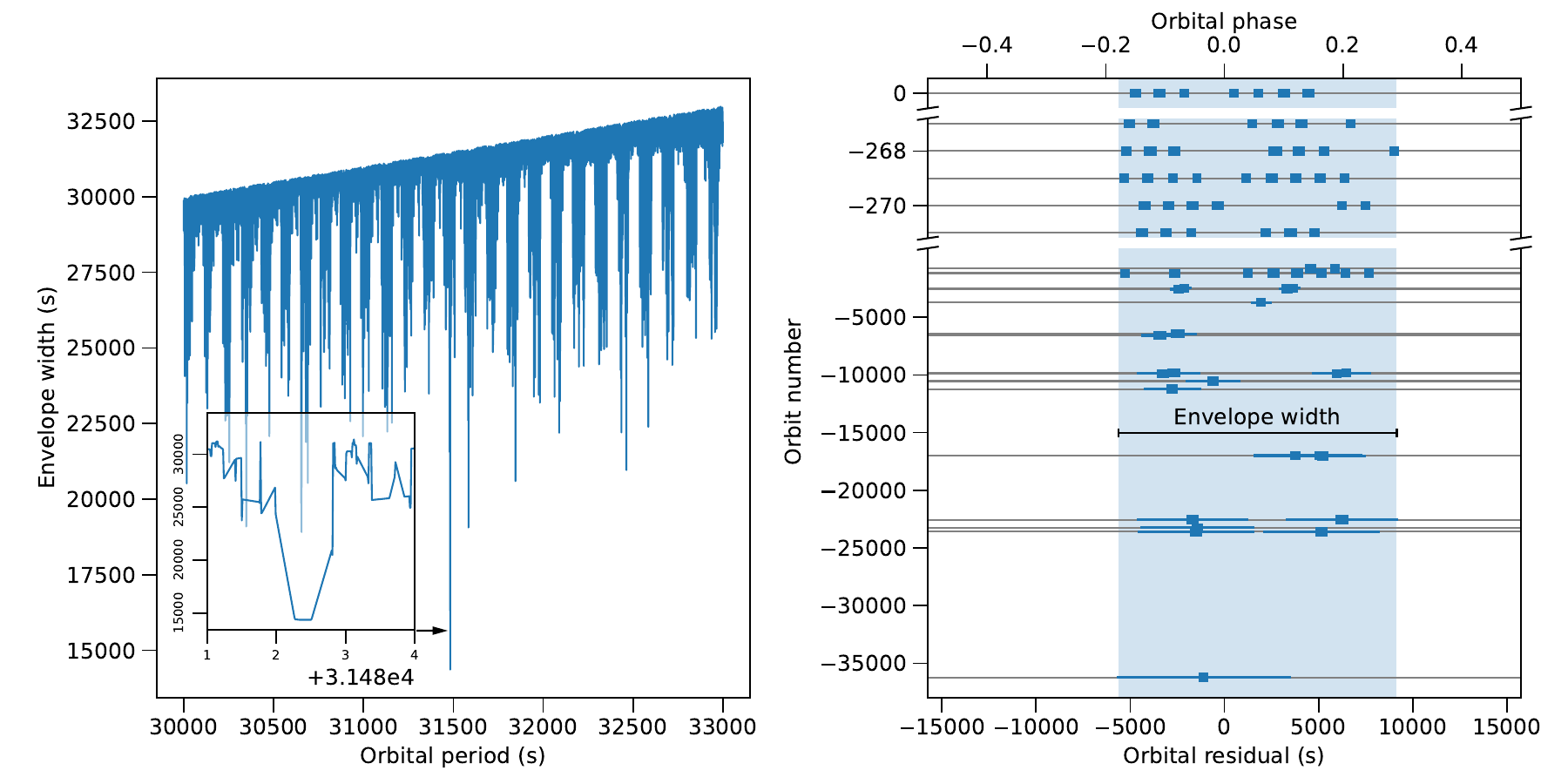}
\caption{Left: Envelope width for trial orbital periods. Right: Detected pulses folded on the orbital period. The y axis is the orbital period number since the most recent detection (negative means backward in time). The widths of the blue rectangles are the pulse widths. The blue lines are the uncertainties in residual given the uncertainty in period. The envelope width is measured from the leftmost pulse edge to the rightmost pulse edge.\label{fig:abacus}}
\end{figure}

To measure the orbital period, the sparseness of the data and the nature of the periodicity makes standard techniques like Fourier analysis and phase dispersion measure unreliable.
\autoref{fig:abacus} shows all pulses detected of \gpmlpt{} folded on the orbital period, including those found in historical data, with the oldest detection in 1988. See \cite{2023Natur.619..487H} for details of the historical detections. The y axis is the number of orbits since the most recent detection on UTC 09/11/2024 06:17:30, and the x axis is the residual, with orbital phase 0 chosen to be between the two pulse groups. The orbital envelope width is measured from the leftmost pulse edge to the rightmost pulse edge. If we assume that the historical pulses arrived in the same range of orbital phase as the recent pulses, then for a convincing timing solution all pulses should lie in a central $\sim$12000 s column on \autoref{fig:abacus}.

Pulses were determined to be the regions of the light curves for which flux $>$ 10 mJy and flux $>$ 6 $\times$ RMS. The RMS of lightcurves was calculated only for the regions of the lightcurves for which flux $<$ 10 mJy to avoid including the pulses in the noise measurement.
Trial periods were searched in steps of 0.01 s, and for each period the envelope width is shown in \autoref{fig:abacus}. Envelope widths greater than half of the period are meaningless since the measurement depends on the choice of phase centre. There is only one minimum envelope width less than half the period at \gpmlptT{}. The period uncertainty is the width of the plateau at the minimum in the zoomed cutout. The uncertainty bars are the residual uncertainties given the uncertainty in period. All trial periods outside of the uncertainty range result in some of the historical pulses outside of the envelope of the most recent observations.

\subsection{Geometric model}

To describe the system geometry, we place the WD at the origin with $\hat{z}$ as the WD spin axis. The orbital axis is inclined about $\hat{y}$ by angle $i$, and the WD magnetic moment $\mu$ is inclined from $\hat{z}$ by angle $\alpha$. \autoref{fig:scale_diagram} shows a diagram of the binary system model. The unit vector of the WD magnetic moment in Cartesian coordinates is
\begin{equation}\label{eqn:beam_posvec}
    \hat{\mu} = (\cos\phi\sin\alpha, ~\sin\phi\sin\alpha, ~\cos\alpha)
\end{equation}
The unit vector from the origin to the MD is
\begin{equation}\label{eqn:md_posvec}
    \hat{r}_\text{MD} = (\cos{i}\cos\phi_\text{orb}, ~\sin\phi_\text{orb}, ~-\sin{i}\cos\phi_\text{orb})
\end{equation}
The unit vector from the origin to the Earth is
\begin{equation}\label{eqn:earth_posvec}
    \hat{r}_\text{E} = (\sin\zeta\cos\phi_0, ~\sin\zeta\sin\phi_0, ~\cos\zeta)
\end{equation}
The angle $\beta_\text{MD}$ between the MD and the WD beam is then
\begin{equation}\begin{aligned}\label{eqn:bead_ang}
    \hat{\mu}\cdot\hat{r}_\text{MD} &= \cos\beta_\text{MD} \\
    \beta_\text{MD}(\phi_\text{orb},\phi) &= \arccos\left(\cos\phi\sin\alpha\cos{i}\cos\phi_\text{orb} + \sin\phi\sin\alpha\sin\phi_\text{orb} - \cos\alpha\sin{i}\cos\phi_\text{orb}\right)
\end{aligned}\end{equation}
and the angle $\beta$ between the Earth and the WD beam is
\begin{equation}\begin{aligned}\label{eqn:earth_beam_ang}
    \hat{r}_\text{E}\cdot\hat{\mu} &= \cos\beta \\
    \beta(\phi) &= \arccos\left(\sin\zeta\sin\alpha\cos(\phi_0-\phi) + \cos\zeta\cos\alpha\right)
\end{aligned}\end{equation}
We assume that the radio beam has a Gaussian cross-section. This shape of course cannot reproduce the intricate structure, but given the variability a simple shape was decided to be the best approximation. We set the Gaussian's standard deviation to be 1/5$^\text{th}$ of the full beam width $W_\text{spin}$. Then, the fractional intensity observed on Earth is
\begin{equation}\label{eqn:beam}
    f_\text{spin}(\beta) = \exp\frac{-\beta^2}{2(W_\text{spin}/5)^2}
\end{equation}
where $W_\text{spin}$ is the opening angle in radians.
We also assume that the interaction with the MD modulates the intensity with a Gaussian profile with a maximum at $\beta_\text{MD} = 0$ when the beam is pointed towards the MD. The orbital modulation function is
\begin{equation}\label{eqn:modulation}
    f_\text{orb}(\beta_\text{MD}) = \exp\frac{-\beta_\text{MD}^2}{2(W_\text{orb}/5)^2}
\end{equation}
The predicted intensity $I_\text{pred}$ observed on Earth is the product $f_\text{spin}(\beta) f_\text{orb}(\beta_\text{MD})$, but to account for either pole producing emission we add the same product offset by $\pi$ radians. Thus,
\begin{equation}\label{eqn:flux}
    I_\text{pred}(\beta,\beta_\text{MD}) = C [f_\text{spin}(\beta) f_\text{orb}(\beta_\text{MD}) + f_\text{spin}(\beta-\pi) f_\text{orb}(\beta_\text{MD}-\pi)]
\end{equation}
where $C$ is a coefficient with units of Jy.

\subsection{Model fitting}

We used the ensemble Markov chain Monte Carlo (MCMC) sampler \verb|emcee| \cite{2013PASP..125..306F}. We define the likelihood $P_\text{fit}$ that the model fits the data as
\begin{equation}{\label{eqn:pfit}}
    \ln P_\text{fit} = -\frac{1}{2} \sum_{k=1}^N \left[ \left(\frac{I_\text{obs}^{(k)} - I_\text{pred}^{(k)}}{\sigma_k}\right)^2 + \ln(2\pi\sigma_k^2) \right]
\end{equation}
where $N$ is the number of data points, $I_\text{obs}^{(k)}$ and $I_\text{pred}^{(k)}$ are the observed and predicted intensities for the $k^\text{th}$ data point respectively, and $\sigma_k$ is the $k^\text{th}$ uncertainty.
The angles $\phi_\text{orb}^{(k)}$ and $\phi^{(k)}$ were calculated for each data point using the timing residuals $\tau$ for the spin and orbital periods
\begin{equation}\label{eqn:calc_res}
    \tau(t,p) = t - t_0 - \left\lfloor\frac{t-t_0}{p}+\frac{1}{2}\right\rfloor p
\end{equation}
where $p$ is the respective period and $t_0$ is the reference time. Then,
\begin{equation}\label{eqn:calc_alpha}
    \phi_\text{orb}^{(k)} = \frac{2\pi}{P_\text{orb}}\tau(t_k,P_\text{orb}) + \phi_\text{orb}^{(0)}
\end{equation}
\begin{equation}\label{eqn:calc_beta}
    \phi^{(k)} = \frac{2\pi}{P}\tau(t_k,P) + \phi^{(0)}
\end{equation}
where $\phi_\text{orb}^{(0)}$ and $\phi^{(0)}$ are the $\phi_\text{orb}$ and $\phi$ angles at the reference time $t_0$.

\begin{figure}[H]
\centering
\includegraphics[width=\linewidth]{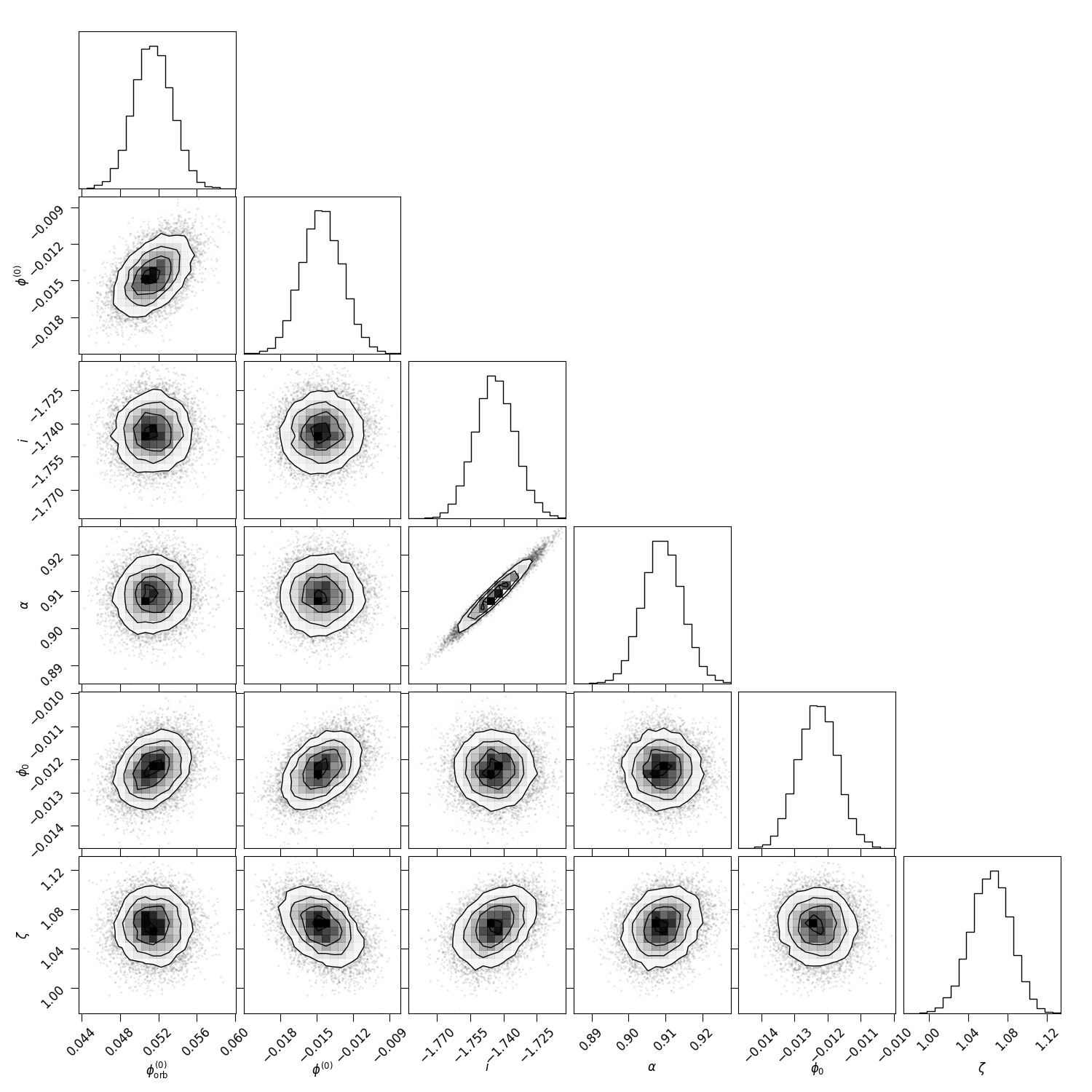}
\caption{MCMC corner plot for \gpmlpt{}. The axis units are radians. \label{fig:mcmc_corner_plot}}
\end{figure}

The beam and modulation widths $W_\text{spin}$ and $W_\text{orb}$ were not fit using MCMC because it tended to find unreasonable local $P_\text{fit}$ maxima (such as $W_\text{spin} = W_\text{orb} = 0$) because the real pulse profiles are so non-Gaussian. Rather than an explicit description of the beam profile, the Gaussian functions may be thought of as a weighting for how well the predicted pulse region on the spin-orbit dynamic pulse profile agrees with the region where pulses were observed.  Instead of fitting the widths, we calculate the minimum $W_\text{spin}$ and $W_\text{orb}$ such that all observed pulses would be observable given the other parameters:
\begin{equation}\label{eqn:beam_width_calc}
    W_\text{spin} = \max_{k\in J} \left\lbrace \beta (\phi^{(k)}) \right\rbrace
\end{equation}
\begin{equation}\label{eqn:mod_width_calc}
    W_\text{orb} = \min_{k\in J} \left\lbrace \beta_\text{MD}(\phi_\text{orb}^{(k)},\phi^{(k)}) \right\rbrace
\end{equation}
where $J$ is the set of data indices which contain pulses according to the conditions in \nameref{subsec:timing_analysis}.
The coefficient $C$ was also not fit using MCMC because it can be simply calculated analytically by solving for the roots of the derivative of \autoref{eqn:pfit} which gives
\begin{equation}
    C = \frac{\sum_{k=1}^N {I_\text{pred*}^{(k)} I_\text{obs}^{(k)}}/{\sigma_k^2}}{\sum_{k=1}^N \left({I_\text{pred*}^{(k)}}/{\sigma_k} \right)^2}
\end{equation}
where $I_\text{pred*} = I_\text{pred}/C$. This way, the model parameters are reduced to 6 $(\phi_\text{orb}^{(0)}, \phi^{(0)}, i, \alpha, \phi_0, \zeta)$, and the parameter space is quite stable as shown by \autoref{fig:mcmc_corner_plot}.

\autoref{fig:mcmc_corner_plot} is a corner plot for the fitting of \gpmlpt{} folded on $P_2$. There is a linear relationship between $i$ and $\alpha$, but their range is limited by the other parameters. This degeneracy becomes significant if there is only one pulse group, such as for \jlpt{}. If we choose instead $P_1$ as the spin period, no good fit can be found because it is impossible for this model to predict a dynamic pulse profile where two pulse groups are on top of each other. We also tried other beat periods, and only $P_2$ produces a good fit. This makes us confident that $P_2$ is the spin period.
It should be noted that reflections along $\hat{x}$ and/or $\hat{z}$ lead to identical dynamic pulse profiles, so those transformations can be applied to get more pleasing angles.

\begin{table}
    \begin{center}
        \begin{tabular}{ccc}
                \hline
                             & Variable        & Value                             \\
                \hline
                Priors       & $P_\text{spin}$ & 1265.2197 $\pm$ 0.0002 s          \\
                             & $P_\text{orb} $ & 31482.4   $\pm$ 0.2 s             \\
                             & $M_\text{WD}  $ & [0.600, 1.200]          M$_\odot$ \\
                             & $M_\text{MD}  $ & [0.140, 0.500]          M$_\odot$ \\
                \hline
                Fitted       & $i            $ &  100.1  $\pm$ 0.6       $^\circ $ \\
                             & $\alpha       $ &   52.1  $\pm$ 0.4       $^\circ $ \\
                             & $\phi_0       $ &  179.29 $\pm$ 0.04      $^\circ $ \\
                             & $\zeta        $ &   61    $\pm$ 2         $^\circ $ \\
                             & $W_\text{spin}$ &   65    $\pm$ 3         $^\circ $ \\
                             & $W_\text{orb} $ &   70    $\pm$ 1         $^\circ $ \\
                \hline
                Calculated   & $A            $ & [1.934, 2.551]          R$_\odot$ \\
                             & $R_\text{MD}  $ & [0.207, 0.574]          R$_\odot$ \\
                             & $R_c          $ & [0.212, 0.267]          R$_\odot$ \\
                             & $R_{rl}       $ & [0.513, 0.784]          R$_\odot$ \\
                             & $R_{lc}       $ & 86.29970  $\pm$ 0.00002 R$_\odot$ \\
                             & $R_{A}        $ & [0.187, 0.828]          R$_\odot$ \\
                \hline
        \end{tabular}
    \end{center}
    \caption{Table of results for \gpmlpt{}. Top are priors for the MCMC fit and calculations. Middle are the results of the MCMC fit, with the uncertainties being the $1\sigma$ credible intervals. Bottom are results which do not depend on the fitted geometry.}
    \label{tab:fit_results}
\end{table}

\begin{table}
    \begin{center}
        \begin{tabular}{ccc}
                \hline
                             & Variable        & Value                     \\
                \hline
                Priors       & $P_\text{spin}$ &   319.3490 $\pm$ 0.0001 s \\
                             & $P_\text{orb} $ & 14525.5588 $\pm$ 0.0001 s \\
                             & $i            $ & [ 53,  65] $^\circ$       \\
                \hline
                Fitted       & $\alpha       $ & [ 42,  70] $^\circ$       \\
                             & $\phi_0       $ & [-48, -17] $^\circ$       \\
                             & $\zeta        $ & [ 39,  73] $^\circ$       \\
                             & $W_\text{spin}$ & [ 20,  60] $^\circ$       \\
                             & $W_\text{orb} $ & [ 67,  90] $^\circ$       \\
                \hline
        \end{tabular}
    \end{center}
    \caption{Table of results for \jlpt{}. Top are priors for the MCMC fit and calculations. Middle are the results of the MCMC fit, with the uncertainties being the $1\sigma$ credible intervals.}
    \label{tab:j1912_results}
\end{table}

\subsection{Alternative geometries}

\begin{figure}[H]
\centering
\includegraphics[width=\linewidth]{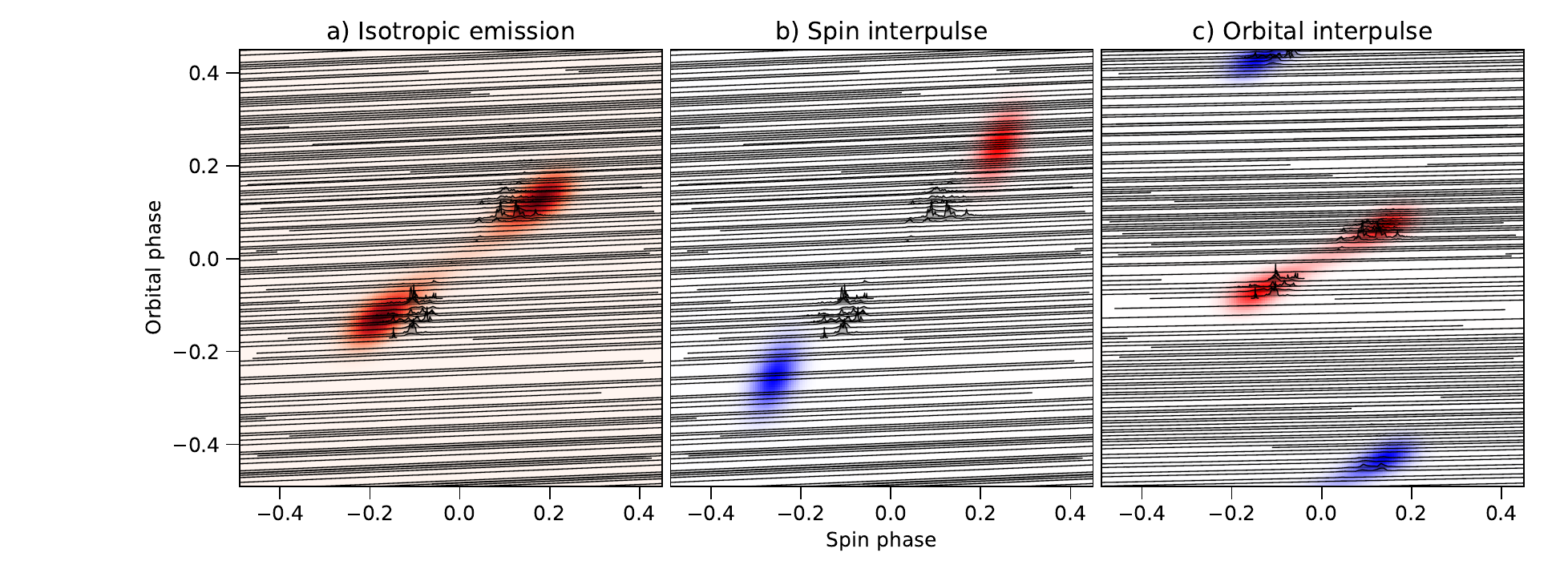}
\caption{Best fits for alternative configurations. White indicates no predicted emission, and red or blue indicate emission from different sources. a) The emission is isotropic. b) The two pulse groups originate from opposite WD poles, indicated by red and blue. c) The emission is 180\textdegree{} bi-directional, regardless of which pole sweeps across the MD. In this case, the orbital period is doubled. Red and blue indicates whether the WD is behind or in front of the MD. \label{fig:alternatives}}
\end{figure}

The three panels of \autoref{fig:alternatives} are alternative interpretations of the geometry which we investigated but ultimately rejected.
\begin{enumerate}[label=\alph*.]
    \item What if the emission is not in the direction of the WD magnetic moment, and the observed flux density is purely a function of $\beta_\text{MD}$, the angle between the WD magnetic moment and the MD? While the predicted pulse group pattern does resemble the observations, no better fit is possible than pictured. This means the emission must be beamed in (or close to) the direction of the WD magnetic moment.

    \item What if the two pulse groups are from opposite poles of the WD? The two pulse groups must be separated by $\sim$0.5 spin phase, so this interpretation was discarded.

    \item What if the emission is 180\textdegree{} bi-directional, regardless of which WD pole sweeps the MD? In this case, we should see emission around orbital phases 0.0 and 0.5, which requires doubling the orbital period to $\sim$18 hours. Although the predicted placement of the pulse groups is correct, there is no qualitative difference between the observed pulses at opposite orbital phases in either total flux density, polarised flux density, or position angle. We therefore favour the $\sim$9 hour orbit.
\end{enumerate}

\subsection{General calculations}\label{subsec:other_calcs}

Using a range of MD masses of $M_\text{MD} \in [0.14, 0.5] M_\odot$ and a range of WD masses of $M_\text{WD} \in [0.6, 1.2] M_\odot$, the orbital separation $A$ assuming a circular orbit is
\begin{equation}\label{eqn:orbital_radius}
    A = \left(\frac{G (M_\text{MD} + M_\text{WD}) P_\text{orb}^2}{4\pi^2}\right)^{\frac{1}{3}} \in [1.93, 2.55] R_\odot
\end{equation}
Approximating the Roche lobe as a sphere, its radius $R_{rl}$ is
\begin{equation}\label{eqn:roche_lobe}
    R_{rl} = A (0.46224) \left( \frac{q}{1+q} \right)^{\frac{1}{3}} \in [0.513, 0.784] R_\odot
\end{equation}
where $q = M_\text{MD} / M_\text{WD}$.
Assuming an MD stellar radius $R_\text{MD}$ of
\begin{equation}\label{eqn:stellar_radius}
    R_\text{MD} = \left(\frac{M_\text{MD}}{M_\odot}\right)^{0.8} R_\odot \in [0.207, 0.574] R_\odot
\end{equation}
The light cylinder radius $R_{lc}$ is the radius at which the angular velocity of the WD matches the speed of light. We find
\begin{equation}\label{eqn:light_cylinder}
    R_{lc} = \frac{c P}{2\pi} = 86.29970 \pm 0.00001  R_\odot
\end{equation}
The corotation radius $R_c$ is the radius at which the orbital period is equal to the spin period. Using \autoref{eqn:orbital_radius} we get
\begin{equation}\label{eqn:corotation}
    R_c = \left(\frac{G (M_\text{MD} + M_\text{WD}) P^2}{4\pi^2}\right)^{\frac{1}{3}} \in [0.212, 0.267] R_\odot
\end{equation}
To calculate the Alfvén radius $R_A$ we assume $\dot{M}_\text{WD}$ = $10^{-14}$\,M$_\odot$ yr$^{-1}$ for a solar-like star \cite{2005ApJ...628L.143W}, a WD magnetic field $B_\text{WD} \sim 10^8$\,G, and a WD radius $R_\text{WD} \in [4000, 9000]$\,km. Then,
\begin{equation}\label{eqn:alfven}
    R_A = \left( \frac{2R_\text{WD}^6 B_\text{WD}^2}{2\dot{M}_\text{MD}\sqrt{G M_\text{WD}}} \right)^{2/7} \in [0.187, 0.828] R_\odot
\end{equation}
The distance $A_\text{WD}$ of the WD from the centre of mass around which the system rotates is
\begin{equation}\label{eqn:dist_from_cm}
    A_\text{WD} = \frac{A M_\text{MD}}{M_\text{MD} + M_\text{WD}} \in [0.246, 1.003] R_\odot
\end{equation}
The maximum difference in light travel time $\Delta t$ from the WD between opposite orbital phases is
\begin{equation}\label{eqn:light_travel_time}
    \Delta t = 2 \frac{A_\text{WD}}{c} \in [1.15, 4.68] s
\end{equation}
where $c$ is the speed of light. The maximum possible time delay in our model is much smaller because we do not view the orbital plane edge-on. We did not correct the timing residuals for this delay because it is of the scale of our time resolution and insignificant compared with the intrinsic arrival time variability of the source.

\subsection{Radio luminosity}\label{subsec:luminosity}

\revised{The brightest pulse from \gpmlpt{} detected had peak $S_\text{1GHz} = 279$\,mJy, but due to the variability in pulse brightness and shape, we use the mean flux density of the brightest pulse over the duty cycle $\bar{S}_\text{1GHz} = 47$\,mJy to estimate the isotropic radio luminosity using the spectral shape from \autoref{eqn:spectrum} as}
\begin{equation}
    L_{4\pi} = 4\pi d^2 \bar{S}_\text{1GHz} \int_0^\infty \left( \frac{\nu}{1\text{GHz}} \right)^\alpha \exp q \left( \log \frac{\nu}{1\text{GHz}} \right)^2 d\nu = 3.5_{-2.7}^{+4.5}\times10^{31}\text{\,erg\,s}^{-1}
\end{equation}
\revised{where $d = 5.7 \pm 2.9$\,kpc is the DM distance \cite{2023Natur.619..487H}. We make no assumptions regarding the beam opening angle's frequency dependence, and estimate the luminosity given the fitted beam opening angle $W_\text{spin}$ as}
\begin{equation}
    L_\Omega = L_{4\pi} \sin^2 \frac{W_\text{spin}}{2} = 1.0_{-0.8}^{+1.3}\times 10^{31} \text{\,erg\,s}^{-1}\text{.}
\end{equation}

\subsection{Magnetospheric interaction}\label{subsec:magnetospheric_interaction}

\revised{Recent theoretical work \cite{2025arXiv250909224Y} has examined the unipolar induction and magnetospheric interaction models as two phases of WD -- MD binary LPT progenitor. By considering the orbital separation at which the WD's magnetic field dominates at the MD location, the authors find the unipolar induction can be active up to $P_\text{orb}\sim3$\,hour orbital periods given the WD is magnetic. This would place \gpmlpt{} with $P_\text{orb} \sim 8.75\text{\,hr}$ in the magnetospheric interaction phase.
Using Equation 39 of their work and the above established parameter bounds, we estimate the energy dissipation rate in this phase as}
\begin{equation}\begin{aligned}
    \dot{E} \simeq & 8.6\times10^{31}\text{\,erg\,s}^{-1} \left( \frac{B_\text{WD} R_\text{WD}^3}{10^{34}\text{\,G\,cm}^3} \right) \left( \frac{B_\text{MD} R_\text{MD}^3}{10^{33}\text{\,G\,cm}^3} \right)
    \left( \frac{P_\text{orb}}{100\text{\,min}} \right)^{-3} \left( \frac{M_\text{WD}+M_\text{MD}}{M_\odot} \right)^{-1} \left( \frac{\Delta\Omega}{\Omega} \right) \\
    \in & \left\{ \begin{matrix}
        [0.042, 3.95] \text{ for } B_\text{WD} = 10^6\text{\,G}\\
        [4.20, 395] \text{ for } B_\text{WD} = 10^8\text{\,G}
    \end{matrix} \right|  \times 10^{31} \text{\,erg\,s}^{-1}
\end{aligned}\end{equation}
\revised{where $B_\text{MD} \sim 10^3$\,G is the MD surface magnetic field strength, $\Omega=2\pi/P_\text{orb}$ is the orbital angular velocity, and $\Delta\Omega=|2\pi/P-\Omega|$ is the relative angular velocity between the WD magnetosphere and the orbit. Therefore, even with a weak WD magnetic field, the required luminosity can be reached.}

\subsection{Pulsar-like geometry}

In the neutron star pulsar case the emission site is bounded by the last open field lines of the dipolar magnetic field of the WD at the emission site at distance $r_{em}$ from the WD. Assuming that the emission site is within the orbital radius, $r_{em} \leq A \leq 2.55 R_\odot$, the angle $\theta$ between $\vec{\mu}$ and the emission site boundary is
\begin{equation}
    \theta = \arcsin\sqrt{\frac{r_{em}}{R_{lc}}} \leq 9.898 ^\circ
\end{equation}
For a dipolar magnetic field in a spherical coordinate system,
\begin{equation}\label{eq:bvec}
    \vec{B} \propto 2 \cos \theta \hat{r} + \sin\theta \hat{\theta}
\end{equation}
If the radiation at the emission site is generated tangential to the magnetic field line, then the angle $\rho$ of $\vec{B}$ with respect to $\vec{\mu}$ at the emission site boundary is the half-opening angle of the beam. We calculate the maximum half-opening angle using \autoref{eq:bvec} by
\begin{equation}
    \tan(\rho - \theta) = \frac{\sin\theta}{2\cos\theta} = \frac{1}{2}\tan\theta
\end{equation}
to get $\rho \leq 14.9^\circ$.

\subsection{Spin-down energy}\label{subsec:spindown}

\revised[R1.4,R1.7]{Despite a more accurate model of the pulse times of arrival, we cannot constrain the spin period derivative beyond that which is already published ($|\dot{P}| \lesssim 3.6\times10^{-13}$\,s\,s$^{-1}$)\cite{2023Natur.619..487H} because the period fit quality is dominated by the 36 year lever arm, as well as the width and variability of the pulses. In fact, we cannot even measure the sign of $\dot{P}$.}
\revised{Nevertheless, using the aforementioned limit, we estimate the spin-down energy dissipation rate for neutron star and white dwarf cases as}
\begin{equation}
    \dot{E} = 4\pi^2 I \dot{P} P^{-3}
\end{equation}
\revised{where $I = 2/5 M R$ is the moment of inertia of a filled sphere of mass $M$ and radius $R$. For a neutron star with $M = 1.4 M_\odot$ and $R = 10$\,km we find $\dot{E}_\text{NS} \lesssim 7.8\times10^{24}$\,erg\,s$^{-1}$. For a white dwarf with mass and radius in the previously defined ranges, we find $\dot{E}_\text{WD} \lesssim [0.6, 5.5]\times10^{30}$\,erg\,s$^{-1}$. $\dot{E}_\text{NS}$ cannot account for the observed luminosity with any reasonable beam opening angle, but $\dot{E}_\text{WD}$ does overlap $L_\Omega$ due to the larger WD moment of inertia. Therefore, in principle there might be enough spin-down energy in the WD alone to power the emission, but the tenuous overlap between the $\dot{E}_\text{WD}$ and $L_\Omega$ limits suggests the additional angular momentum of the orbit is necessary. All of the mechanisms discussed are in some way powered by the relative magnetospheric motions, so the spin-down energy budget is likely satisfied.}

\section*{Acknowledgements}
This research is supported by an Australian Government Research Training Program (RTP) Scholarship \href{doi.org/10.82133/C42F-K22}.
N.H.-W. is the recipient of an Australian Research Council Future Fellowship (project number FT190100231).
N.R. is supported by the European Research Council (ERC) via the Consolidator Grant “MAGNESIA” (No. 817661) and the Proof of Concept ``DeepSpacePulse" (No. 101189496), by the Catalan grant SGR2021-01269
(PI: Graber/Rea), the Spanish grant ID2023-153099NA-I00 (PI: Coti Zelati), and by the program Unidad de Excelencia Maria de Maeztu CEX2020-001058-M.

This scientific work uses data obtained from Inyarrimanha Ilgari Bundara, the CSIRO Murchison Radio-astronomy Observatory. We acknowledge the Wajarri Yamaji People as the Traditional Owners and native title holders of the Observatory site. CSIRO’s ASKAP radio telescope is part of the Australia Telescope National Facility (\url{https://ror.org/05qajvd42}). Operation of ASKAP is funded by the Australian Government with support from the National Collaborative Research Infrastructure Strategy. ASKAP uses the resources of the Pawsey Supercomputing Research Centre. Establishment of ASKAP, Inyarrimanha Ilgari Bundara, the CSIRO Murchison Radio-astronomy Observatory and the Pawsey Supercomputing Research Centre are initiatives of the Australian Government, with support from the Government of Western Australia and the Science and Industry Endowment Fund.

The MeerKAT telescope is operated by the South African Radio Astronomy Observatory, which is a facility of the National Research Foundation, an agency of the Department of Science and Innovation.
The National Radio Astronomy Observatory is a facility of the National Science Foundation operated under cooperative agreement by Associated Universities, Inc.

We acknowledge Ben Stappers, Ewan Barr, and Yunpeng Men for contributing to the MeerKAT proposal. We acknowledge Scott Hyman, Tracy Clarke, and Simona Giacintucci for contributing to the VLA proposal. We thank Ingrid Pelisoli and Simone Scaringi for useful discussions on white dwarf evolution.

\section*{Author contributions}

Cs.H performed the orbital timing, spin/beat period analysis, polarisation analysis, developed the geometric model, and contributed to the discussion in the manuscript.
N.R contributed to the formulation of the model and its physical interpretation, and contributed to the discussion in the manuscript.
N.H.-W led the proposals for the radio data toward \gpmlpt{}, imaged and produced dynamic spectra from the radio data used in this work, and contributed to the formulation of the model and discussion in the manuscript.
S.M contributed to the formulation of the model, and contributed to the discussion in the manuscript.
R.P generated the observing scripts and calibrated the data for the VLA observations.
E.L processed the ASKAP data and contributed to the polarisation analysis.

\section*{Competing interests}
The authors declare no competing interests.

\section*{Data availability}
Data supporting the findings of this study are available from the corresponding author upon request.

\section*{Code availability}
The code used for the geometric model and MCMC fitting is available at\\\url{https://github.com/CsanadHorvath/GeometricLPT.git}.\\
An interactive web visualisation is available at \\\url{https://www.desmos.com/3d/6bisspujwe}.


%


\end{document}